\documentclass{pasj00}
\begin{document}
\SetRunningHead{H. Uchiyama et al.}
{Distribution of K-shell Lines along the Galactic Plane}

\title{The K-shell Line Distribution of Heavy Elements along the Galactic Plane Observed with Suzaku}
\author{Hideki \textsc{Uchiyama},\altaffilmark{1} Masayoshi \textsc{Nobukawa},\altaffilmark{2,3} Takeshi Go \textsc{Tsuru},\altaffilmark{3} and Katsuji \textsc{Koyama}\altaffilmark{3,4}}
\altaffiltext{1}{Department of Physics, Graduate School of Science, The University of Tokyo, 7-3-1 Hongo, Bunkyo-ku, Tokyo 113-0033}
\altaffiltext{2}{The Hakubi Center for Advanced Research, Kyoto University, Yoshida-Ushinomiya-cho, Kyoto 606-8302}
\altaffiltext{3}{Department of Physics, Graduate school of Science, Kyoto University, Oiwake-cho, Kitashirakawa, Kyoto 606-8502}
\altaffiltext{4}{Department of Earth and Space Science, Graduate School of Science, Osaka University, 1-1 Machikaneyama, Toyonaka, Osaka 560-0043}
\email{uchiyama@juno.phys.s.u-tokyo.ac.jp}


\maketitle

\begin{abstract}
We report the global distribution of the intensities of the	K-shell lines 
from the He-like and H-like ions of S, Ar, Ca and Fe along the Galactic plane. 
From the profiles, we clearly separate the Galactic center X-ray 
emission (GCXE) and the Galactic ridge X-ray emission (GRXE). The intensity profiles 
of the He-like K$\alpha$ lines of S, Ar, Ca and Fe along the Galactic plane are approximately 
similar with each other, while not for the H-like Ly$\alpha$ lines. In particular, the profiles of H-like Ly$\alpha$ of S 
and Fe show remarkable  contrast; a large excess of Fe and almost no excess of S lines in the 
GCXE compared to the GRXE. 
Although the prominent K-shell lines are represented by 
$\sim$1 keV and $\sim$7 keV temperature plasmas, these two temperatures are not equal between the GCXE and GRXE. 
In fact, the spectral analysis of the GCXE and GRXE revealed that the $\sim$1 keV plasma in the GCXE has lower 
temperature than that in the GRXE, and vice versa for the $\sim$7 keV plasma.

\end{abstract}

\section{Introduction}
The HEAO-1 satellite discovered unresolved X-ray emissions from the inner disk of the Milky Way 
\citep{Wo82}, which were designated as the Galactic ridge X-ray emission (GRXE). The EXOSAT satellite found 
that the GRXE was extended by about $\pm\timeform{60D}$ in the longitude and $\pm\timeform{1D}$--
$\timeform{2D}$ in the latitude \citep{Wa85}. 
The Tenma satellite discovered a strong emission line at the energy of about 6.7 keV 
(the 6.7-keV line) in the GRXE \citep{Ko86}.	
The Ginga satellite made survey observations along the Galactic plane with the 6.7-keV line, and 
found a prominent peak at the Galactic center extending about \timeform{1D} 
(Galactic center X-ray emission, GCXE; \cite{Ko89}, \cite{Ya90}, \cite{Ya93}). Subsequently, 
the 6.7-keV line was found to consist of three lines at 6.4, 6.7 and 7.0 keV, which are 
K$\alpha$ lines of neutral K$\alpha$, He-like K$\alpha$, and H-like Ly$\alpha$ of irons, respectively 
(\cite{Ko96}, \cite{Eb08}).

The strong He-like K$\alpha$ and H-like Ly$\alpha$ lines of iron indicate the presence of hot plasma
with a temperature of $kT\sim$5--10 keV.  Two possibilities of the origin of the 
hot plasma have been proposed by many authors.  One is true diffuse plasma  which fills in the Galactic plane.  
A problem of this scenario is the energy source. The total energy of the plasma is 
estimated to be $\sim10^{56}$ erg.  Such a high-temperature plasma cannot be 
gravitationally bounded in the Galactic plane.  
The escape time scale is estimated to be a few times $10^5$ yr. 
Thus huge energy input ($\sim$0.1 supernova yr$^{-1}$) is required 
to keep the plasma  (e.g. Koyama et al. 1986a).  
The other possibility is that the GCXE and GRXE are superposition of unresolved faint
X-ray point sources \citep{Re09}.  However, point sources with iron lines,
like cataclysmic variables and active binaries, cannot explain large equivalent widths
of the iron lines in the GCXE and GRXE (e.g. \cite{Ya09}).  Also the spatial
distribution of the He-like K$\alpha$ line of iron shows a larger excess than the stellar mass distribution
at the Galactic center region \citep{Uc11}.  True origin would be a mixture of diffuse plasma and point sources,
but the mixing ratios are unclear both for the GRXE and GCXE (see \cite{Mu04}, \cite{Eb05}, \cite{Ko09}, \cite{Re07GC}, \cite{Re09}).

Spectral and spatial structures of the GCXE and GRXE plasmas are key information to reveal the origins.
\citet{Ka97} found that the GRXE consists of two plasmas of about 
0.8 and 7 keV temperatures. Similar two- or multi-temperature structures were revealed for the GCXE 
(\cite{Mu04}, \cite{Ry09}, \cite{No10}) and the intermediate region between the GCXE and GRXE \citep{Yua12}. 
The intensity ratio of the He-like and H-like lines from the same atom gives 
the plasma temperature responsible for these line emissions. 
Thus the two plasmas structures in the GCXE and GRXE, 
in the temperature and mixing ratio, would be resolved by the line intensities from many elements and their 
ratios of the same atom. This paper reports the intensity profiles along the Galactic plane in the He-like 
K$\alpha$ and H-like Ly$\alpha$ for S, Ar, Ca and Fe for the first time. 
Based on the profiles, we discuss possible origins for the GCXE and GRXE.
In this paper, uncertainties are quoted at the 1$\sigma$ confidence range unless otherwise stated.

\section{Observations and Data Reduction}\label{ch:obs}

The observations were made with the X-ray Imaging Spectrometer (XIS; \cite{Ko07a}) 
on the focal plane of the X-ray telescope (XRT; \cite{Se07}) onboard the Suzaku satellite \citep{Mi07}. 
The XIS contains four sets of X-ray CCD camera systems (XIS\,0, 1, 2, and 3).
XIS\,0, 2, and 3 have front-illuminated (FI) CCDs, while 
XIS\,1 has a back-illuminated (BI) CCD. 
The XIS\,2 has been out of function since November 2006.
Although the CCD chips were significantly degraded by 
on-orbit particle radiation, the XIS performance had been calibrated well with the checker-flag 
charge-injection method (\cite{Na08}, \cite{Oz09}) until October 2006, and since then, it has been restored with 
the spaced-row charge-injection method (\cite{Pr08}, \cite{Uc09}). 
As a result, the energy resolutions were 
kept within 130--180 eV (FWHM) at 5.9 keV during the observations.

In this paper, we included the data of the series of the Key and Large projects and many ordinary 
proposals targeted on the Galactic center region. 
The early reports on the results from these observations are found in the published papers (e.g. \cite{Ko07b}, \cite{Yua12}). 
As for the GRXE, 
one can refer to the papers by \citet{Eb08} and \citet{Ya09}.
In addition to these data, we collected all the archival Suzaku observations targeted 
on the Galactic plane of $|b|<\timeform{5D}$ between August  2005 and March 2011.

We downloaded all the archival data of $|b|<\timeform{5D}$ from ISAS DARTS\footnote{http://darts.jaxa.jp/}, then  the data reductions were made uniformly on all the data-sets including previously published data-sets.  We used the pipeline-processed cleaned data provided by the DARTS site, and made images of the respective data in the 0.5--4 keV and 4--10 keV bands.
Based on the X-ray images, we selected the fields with no bright point or diffuse sources. The selected data are listed  in table \ref{tab:obslog}.
In the following analyses (next section), the raw X-ray spectra including the cosmic X-ray background (CXB) were made from the central area with a circle of \timeform{8'.9} radius by subtracting the NXB with the same COR of the source data. The non X-ray background (NXB) spectra sorted by the cut-off rigidity (COR) were made
from the many data sets pointing to the night earth using {\tt xisnxbgen} \citep{Ta08}

Using {\tt xissimarfgen} \citep{Is07} and {\tt xisrmfgen}, the effective area of the XRT 
and the response of the XIS  were calculated, and ancillary response files 
and redistribution matrix files were made for respective observations.
Since the responses of the FIs are almost the same, we merged the FI spectra. 
The BI data were not used because the NXB flux is high above the Fe line bands.

%%%%%%%%%%%%
\renewcommand{\arraystretch}{0.9} 
\begin{table*}
 \caption{Observation data list. } \label{tab:obslog}
\scalebox{0.7}{\begin{minipage}{\textheight}
 \begin{tabular}{cccccccccc}
 \hline
 \hline
OBSID& \multicolumn{4}{c}{Pointing direction} & \multicolumn{2}{c}{Observation} &  Exp. & Object name & Spectra$^*$\\
& $\alpha_{\rm 2000.0}(\timeform{D})$ & $\delta_{\rm 2000.0}(\timeform{D})$ & $l(\timeform{D})$ & $b(\timeform{D})$ & Start (UT)  & End (UT) &  (ks)  && \\
\hline
100028020 & 243.66 & -51.18 & -28.00 & -0.15 & 2005-09-18T22:47:36 & 2005-09-19T11:58:41 & 19.3 & HESS J1616-508 BGD1 & GRXE \\
100028030 & 244.46 & -50.69 & -27.30 & -0.15 & 2005-09-20T19:40:17 & 2005-09-21T07:29:24 & 21.9 & HESS J1616-508 BGD2 & \\
100027020 & 266.30 & -29.17 & -0.25 & -0.05 & 2005-09-24T14:17:17 & 2005-09-25T17:27:19 & 42.8 & Sgr A west & GCXE \\
100026020 & 257.38 & -38.82 & -12.37 & 0.71 & 2005-09-25T19:11:40 & 2005-09-26T15:42:09 & 34.9 & RXJ1713-3946-BKGD1 & GRXE \\
100026030 & 257.27 & -41.04 & -14.20 & -0.54 & 2005-09-28T07:09:13 & 2005-09-29T04:25:24 & 37.5 & RXJ1713-3946-BKGD2 & GRXE \\
100037010 & 266.30 & -29.17 & -0.25 & -0.05 & 2005-09-29T04:35:41 & 2005-09-30T04:29:19 & 43.7 & Sgr A west & GCXE \\
500009010 & 281.00 & -4.07 & 28.46 & -0.20 & 2005-10-28T02:40:08 & 2005-10-30T21:30:15 & 93.3 & GALACTIC RIDGE & \\
500019010 & 265.79 & -29.89 & -1.09 & -0.04 & 2006-02-23T10:51:11 & 2006-02-23T20:02:19 & 13.3 & SGR C BGD & \\
500008010 & 270.96 & -22.02 & 8.04 & -0.05 & 2006-04-07T11:49:16 & 2006-04-08T10:54:18 & 40.7 & HESS J1804-216 BGD & \\
500025010 & 9.24 & 64.24 & 121.36 & 1.41 & 2006-06-29T15:42:01 & 2006-06-30T18:15:24 & 51.0 & TYCHO SNR HXD BKGD & \\
501043010 & 242.01 & -52.44 & -29.60 & -0.38 & 2006-09-16T11:02:03 & 2006-09-17T07:14:14 & 43.6 & HESS J1614-518 BG & GRXE \\
501009010 & 266.19 & -28.91 & -0.07 & 0.18 & 2006-09-29T21:26:07 & 2006-10-01T06:55:19 & 51.2 & GC SOUTH BGD & GCXE \\
501049010 & 265.38 & -29.75 & -1.17 & 0.33 & 2006-10-08T10:22:40 & 2006-10-09T02:19:24 & 19.6 & GALACTIC CENTER & \\
501051010 & 265.70 & -29.93 & -1.17 & -0.00 & 2006-10-09T13:40:09 & 2006-10-10T06:44:24 & 21.9 & GALACTIC CENTER & \\
501052010 & 265.50 & -30.21 & -1.50 & 0.00 & 2006-10-10T06:45:09 & 2006-10-10T21:18:14 & 19.3 & GALACTIC CENTER & \\
501053010 & 265.30 & -30.50 & -1.83 & -0.00 & 2006-10-10T21:18:59 & 2006-10-11T10:06:14 & 21.9 & GALACTIC CENTER & \\
501057010 & 266.03 & -30.11 & -1.17 & -0.33 & 2006-10-11T10:07:27 & 2006-10-12T03:28:14 & 20.5 & GALACTIC CENTER & \\
500009020 & 281.00 & -4.07 & 28.46 & -0.20 & 2006-10-15T02:15:12 & 2006-10-17T19:32:19 & 98.9 & GALACTIC RIDGE & \\
501046010 & 265.98 & -28.90 & -0.17 & 0.33 & 2007-03-10T15:03:10 & 2007-03-11T03:55:14 & 25.2 & GALACTIC CENTER & GCXE \\
501047010 & 265.78 & -29.19 & -0.50 & 0.33 & 2007-03-11T03:55:59 & 2007-03-11T19:04:14 & 25.6 & GALACTIC CENTER GC2 & GCXE \\
501048010 & 265.58 & -29.47 & -0.83 & 0.33 & 2007-03-11T19:04:59 & 2007-03-12T08:09:14 & 27.5 & GALACTIC CENTER GC3 & GCXE \\
502021010 & 113.33 & -19.53 & -125.00 & 0.00 & 2007-04-22T20:39:20 & 2007-04-25T10:04:24 & 89.5 & ANTICENTER & \\
502022010 & 266.81 & -28.88 & 0.23 & -0.27 & 2007-08-31T12:33:33 & 2007-09-03T19:00:25 & 134.8 & (L,B)=(0.25,-0.27) & GCXE \\
502053010 & 275.64 & -15.58 & 15.82 & -0.84 & 2007-10-07T02:16:29 & 2007-10-08T18:10:14 & 71.5 & M17 EAST BKG & GRXE \\
502002010 & 267.16 & -29.14 & 0.17 & -0.67 & 2007-10-09T16:40:54 & 2007-10-10T03:40:24 & 23.2 & GC14 & GCXE \\
502003010 & 266.96 & -29.42 & -0.17 & -0.67 & 2007-10-10T03:41:13 & 2007-10-10T15:20:24 & 21.5 & GC15 & GCXE \\
502004010 & 267.48 & -29.31 & 0.17 & -1.00 & 2007-10-10T15:21:17 & 2007-10-11T01:00:24 & 19.9 & GC16 & GCXE \\
502005010 & 267.29 & -29.60 & -0.17 & -1.00 & 2007-10-11T01:01:17 & 2007-10-11T11:32:20 & 20.6 & GC17 & GCXE \\
502006010 & 266.18 & -28.62 & 0.17 & 0.33 & 2007-10-11T11:34:01 & 2007-10-11T23:07:14 & 22.6 & GC18 & GCXE \\
502007010 & 265.86 & -28.45 & 0.17 & 0.67 & 2007-10-11T23:09:15 & 2007-10-12T09:52:14 & 22.0 & GC19 & GCXE \\
502011010 & 266.57 & -28.05 & 0.83 & 0.33 & 2007-10-13T18:51:09 & 2007-10-14T05:30:24 & 23.0 & GC23 & GCXE \\
503006010 & 13.00 & 62.90 & 123.00 & 0.03 & 2008-08-01T00:17:23 & 2008-08-02T20:11:19 & 86.1 & ANTICENTER2 & \\
503007010 & 266.44 & -28.57 & 0.33 & 0.17 & 2008-09-02T10:15:27 & 2008-09-03T22:52:24 & 52.2 & GC LARGEPROJECT1 & GCXE \\
503008010 & 266.78 & -29.13 & 0.00 & -0.38 & 2008-09-03T22:53:29 & 2008-09-05T06:56:19 & 53.7 & GC LARGEPROJECT2 & GCXE \\
503010010 & 266.04 & -29.55 & -0.70 & -0.05 & 2008-09-06T15:56:13 & 2008-09-08T01:39:24 & 53.1 & GC LARGEPROJECT4 & GCXE \\
503013010 & 265.67 & -30.07 & -1.30 & -0.05 & 2008-09-16T00:51:19 & 2008-09-18T04:44:24 & 104.8 & GC LARGEPROJECT7 & \\
503014010 & 265.18 & -30.75 & -2.10 & -0.05 & 2008-09-18T04:46:49 & 2008-09-19T07:32:20 & 55.4 & GC LARGEPROJECT8 & GRXE \\
503015010 & 265.03 & -30.96 & -2.35 & -0.05 & 2008-09-19T07:33:05 & 2008-09-20T09:56:13 & 56.8 & GC LARGEPROJECT9 & GRXE \\
503016010 & 264.87 & -31.17 & -2.60 & -0.05 & 2008-09-22T06:47:49 & 2008-09-23T08:07:17 & 52.2 & GC LARGEPROJECT10 & GRXE \\
503017010 & 264.72 & -31.38 & -2.85 & -0.05 & 2008-09-23T08:08:10 & 2008-09-24T09:21:13 & 51.3 & GC LARGEPROJECT11 & GRXE \\
503018010 & 264.56 & -31.60 & -3.10 & -0.05 & 2008-09-24T09:27:54 & 2008-09-24T22:30:24 & 29.4 & GC LARGEPROJECT12 & \\
503018020 & 264.56 & -31.60 & -3.10 & -0.05 & 2008-10-03T18:05:13 & 2008-10-04T03:42:18 & 12.2 & GC LARGEPROJECT12 & \\
503018030 & 264.56 & -31.60 & -3.10 & -0.05 & 2009-02-19T07:32:01 & 2009-02-19T16:36:24 & 11.9 & GC LARGEPROJECT12 & GRXE \\
503019010 & 264.40 & -31.81 & -3.35 & -0.05 & 2009-02-19T16:37:49 & 2009-02-21T01:15:14 & 52.8 & GC LARGEPROJECT13 & \\
503020010 & 264.24 & -32.02 & -3.60 & -0.05 & 2009-02-21T01:15:55 & 2009-02-22T18:59:14 & 61.1 & GC LARGEPROJECT14 & \\
503072010 & 265.99 & -29.21 & -0.42 & 0.17 & 2009-03-06T02:39:12 & 2009-03-09T02:55:25 & 140.6 & EXTENDED CHIMNEY & GCXE \\
503099010 & 265.18 & -28.53 & -0.22 & 1.13 & 2009-03-10T19:39:08 & 2009-03-11T10:56:14 & 29.7 & GCL1 & GCXE \\
503101010 & 265.27 & -28.86 & -0.45 & 0.88 & 2009-03-16T14:43:17 & 2009-03-17T07:48:24 & 33.9 & GCL3 & GCXE \\
503102010 & 265.35 & -29.19 & -0.70 & 0.65 & 2009-03-17T07:49:09 & 2009-03-18T01:42:24 & 33.7 & GCL4 & GCXE \\
503022010 & 263.25 & -31.95 & -4.00 & 0.70 & 2009-03-18T23:11:24 & 2009-03-19T21:26:24 & 41.3 & LOOP 1 L=356.00 & GRXE \\
503023010 & 263.46 & -31.67 & -3.67 & 0.70 & 2009-03-26T06:37:01 & 2009-03-27T01:53:19 & 31.2 & LOOP 2 L=356.33 & \\
504093010 & 268.30 & -31.66 & -1.50 & -2.80 & 2009-09-17T13:54:31 & 2009-09-19T03:37:14 & 53.2 & GALACTIC BULGE10 & \\
504089010 & 267.55 & -29.60 & -0.05 & -1.20 & 2009-10-09T04:05:59 & 2009-10-10T14:10:06 & 55.3 & GALACTIC BULGE2 & \\
504090010 & 266.69 & -30.84 & -1.50 & -1.20 & 2009-10-13T04:17:20 & 2009-10-14T11:29:06 & 41.3 & GALACTIC BULGE7 & \\
504088010 & 267.22 & -29.37 & -0.00 & -0.83 & 2009-10-14T11:30:56 & 2009-10-15T15:29:19 & 47.2 & GALACTIC BULGE1 & GCXE \\
505080010 & 270.15 & -30.91 & -0.05 & -3.80 & 2010-04-07T17:15:10 & 2010-04-09T21:14:16 & 56.1 & GALACTIC BULGE5 & \\
505083010 & 264.99 & -32.42 & -3.60 & -0.80 & 2010-10-10T14:04:38 & 2010-10-11T21:33:13 & 52.9 & GALACTIC BULGE14 & \\
505078010 & 267.95 & -29.80 & -0.05 & -1.60 & 2011-03-04T19:13:55 & 2011-03-06T05:30:11 & 51.3 & GALACTIC BULGE3 & \\
505084010 & 265.70 & -32.79 & -3.60 & -1.50 & 2011-03-06T05:36:44 & 2011-03-07T13:01:11 & 50.3 & GALACTIC BULGE15 & \\
505079010 & 269.15 & -30.41 & -0.05 & -2.80 & 2011-03-12T06:36:20 & 2011-03-13T10:17:21 & 50.2 & GALACTIC BULGE4 & \\
505087010 & 269.31 & -34.58 & -3.60 & -5.00 & 2011-03-13T10:20:22 & 2011-03-14T12:39:21 & 51.5 & GALACTIC BULGE18 & \\
505082010 & 264.59 & -32.21 & -3.60 & -0.40 & 2011-03-15T13:54:28 & 2011-03-16T14:54:23 & 48.5 & GALACTIC BULGE13 & GRXE \\
\hline
\end{tabular}\\
\footnotemark[$*$] This column shows  whether the data were used to make the mean GCXE or GRXE spectra in figure \ref{fig:model_fit} (see section \ref{ch:sp}).
\\
\end{minipage}}
\end{table*}
 \renewcommand{\arraystretch}{1.0} 

\section{Analyses}
\subsection{The Line Intensities}

We made the NXB-subtracted X-ray spectra of the GCXE and GRXE, and found  many prominent
lines as is shown in figure \ref{fig:Full_Band_w_low_model} for example. 
From the center energies, we identified these lines as the K-shell lines from He- and H-like   
Ne, Mg, Si, S, Ar, Ca and Fe ions. 
We also identified a prominent K-shell line from neutral Fe (Fe \emissiontype{I} K$\alpha$).
According to \citet{Ry09}, especially in the Galactic center region, the foreground emission dominates below the 2 keV band.  
In order to avoid the contamination of the foreground emission and to extract the line intensities from the GCXE and GRXE, 
we separately fitted the spectra in the 2.3--5 and 5--8 keV 
bands with a phenomenological model, an absorbed power-law continuum plus
nine (for the 2.3--5 keV band) or four (for the 5--8 keV band) narrow Gaussians.

\begin{figure*}[hbtp]
		\begin{center}
		 \FigureFile(80mm,60mm){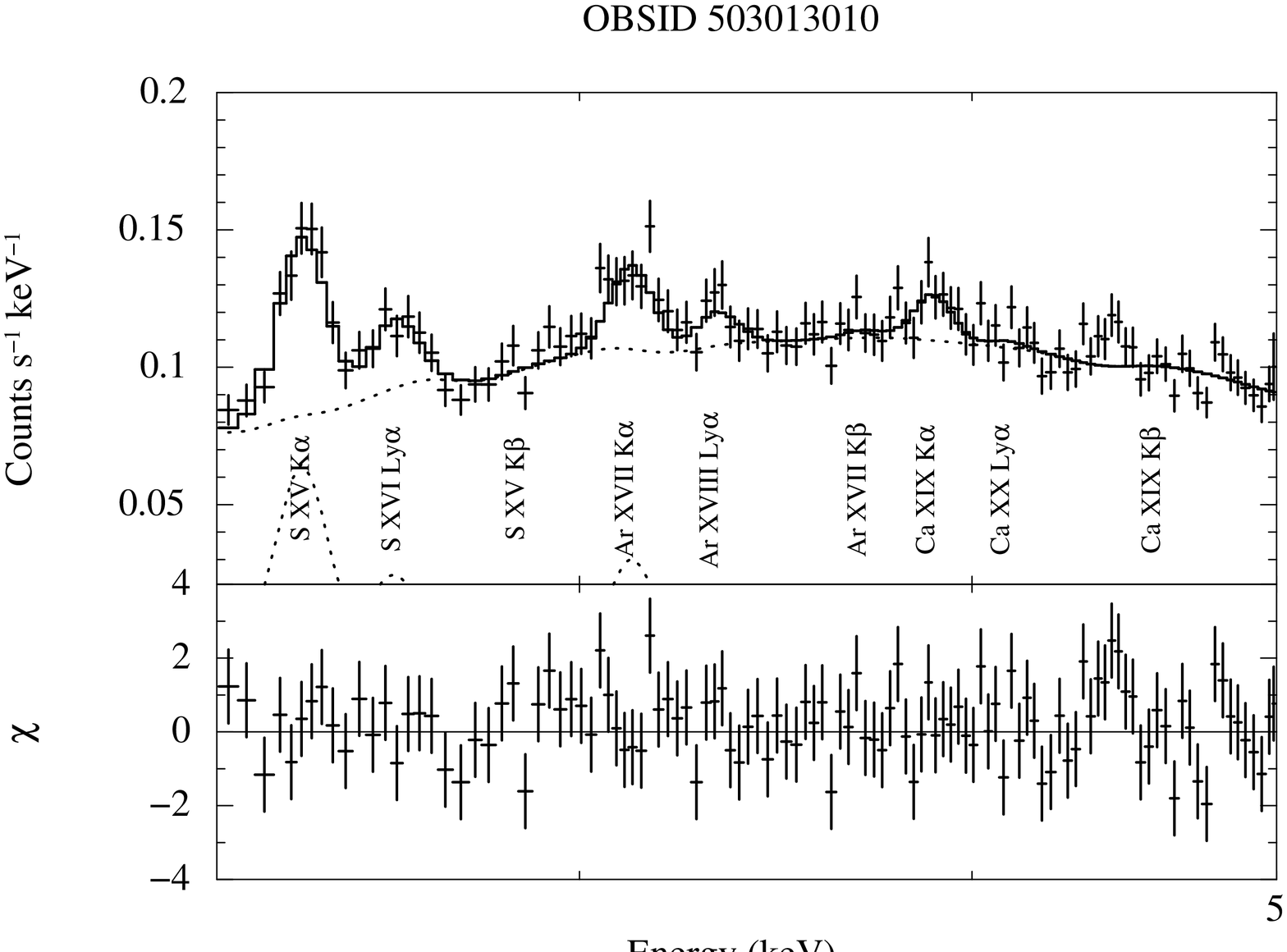}
		 \FigureFile(80mm,60mm){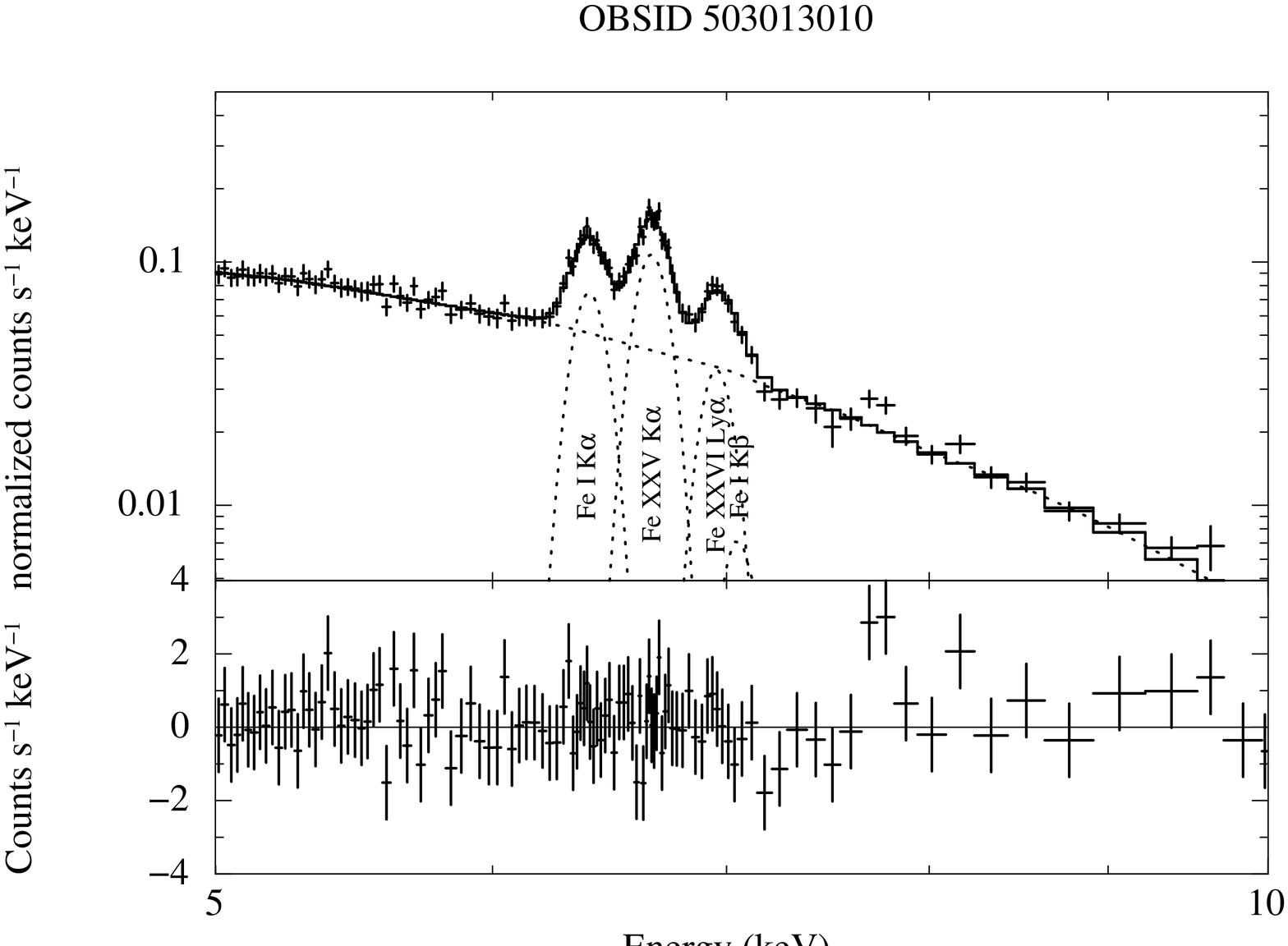}
		\end{center}
	\caption{Left : An example of the spectra with the best-fit phenomenological model in the 2.3--5 keV band (OBSID 503013010). We also show the positions of the lines included in the model. Right : the same as the left panel, but that in the 5--10 keV. }
		\label{fig:SpectraExamples}
\end{figure*}

The line center energies of the Gaussians were fixed to be 
2.45 (S \emissiontype{XV} K$\alpha$), 
2.62 (S \emissiontype{XVI} Ly$\alpha$), 
2.88 (S \emissiontype{XV} K$\beta$), 
3.12 (Ar \emissiontype{XVII} K$\alpha$), 
3.32 (Ar \emissiontype{XVIII} Ly$\alpha$),
3.70 (Ar \emissiontype{XVII} K$\beta$), 
3.89 (Ca \emissiontype{XIX} K$\alpha$), 
4.11 (Ca \emissiontype{XX} Ly$\alpha$) and 
4.59 (Ca \emissiontype{XIX} K$\beta$) for the 2.3--5 keV band, and
6.40 (Fe \emissiontype{I} K$\alpha$), 
6.68 (Fe \emissiontype{XX} K$\alpha$), 
6.97 (Fe \emissiontype{XXI} Ly$\alpha$)  and 
7.06~keV (Fe \emissiontype{I} K$\beta$) for the 5--8 keV band, 
according to  Smith et al.~(2001; APEC model) and \citet{Ka93}.
The intensities of the lines were left free except for Fe \emissiontype{I} K$\beta$, which was fixed to 0.125 times that of the 
Fe \emissiontype{I} K$\alpha$ line according to \citet{Ka93}.
The normalization, photon index, and interstellar absorption in the absorbed power-law 
component were also left free.
The cross section of the photoelectric absorption was obtained from \citet{Mo83}.
Using the phenomenological model, we obtained the intensities of the He-like K$\alpha$ and H-like Ly$\alpha$ lines of S, Ar, Ca, and Fe. We also extracted the intensities in the 2.3--5 keV and 5--8 keV bands, and that of Fe \emissiontype{I} K$\alpha$ line. 
An example of the spectra and the best-fit models is shown in figure \ref{fig:SpectraExamples}.

\begin{figure*}[hbtp]
		\begin{center}
		 \FigureFile(80mm,60mm){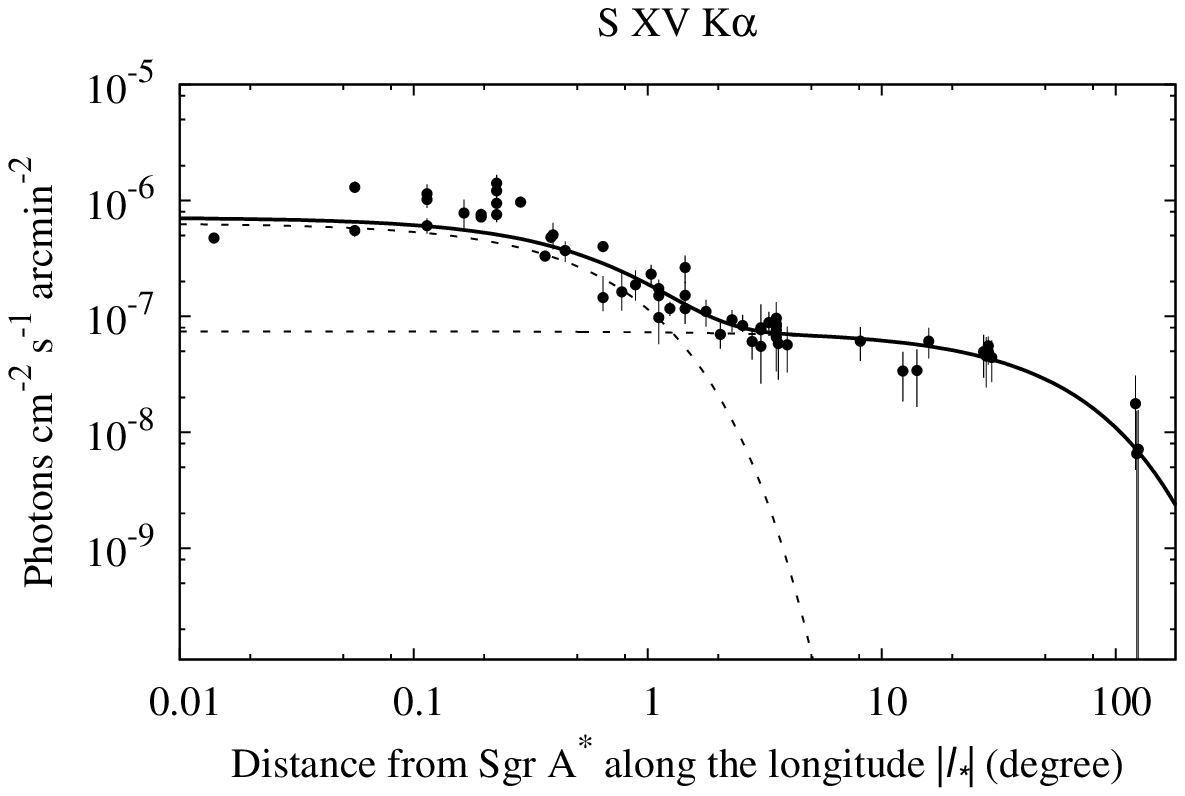}
		 \FigureFile(80mm,60mm){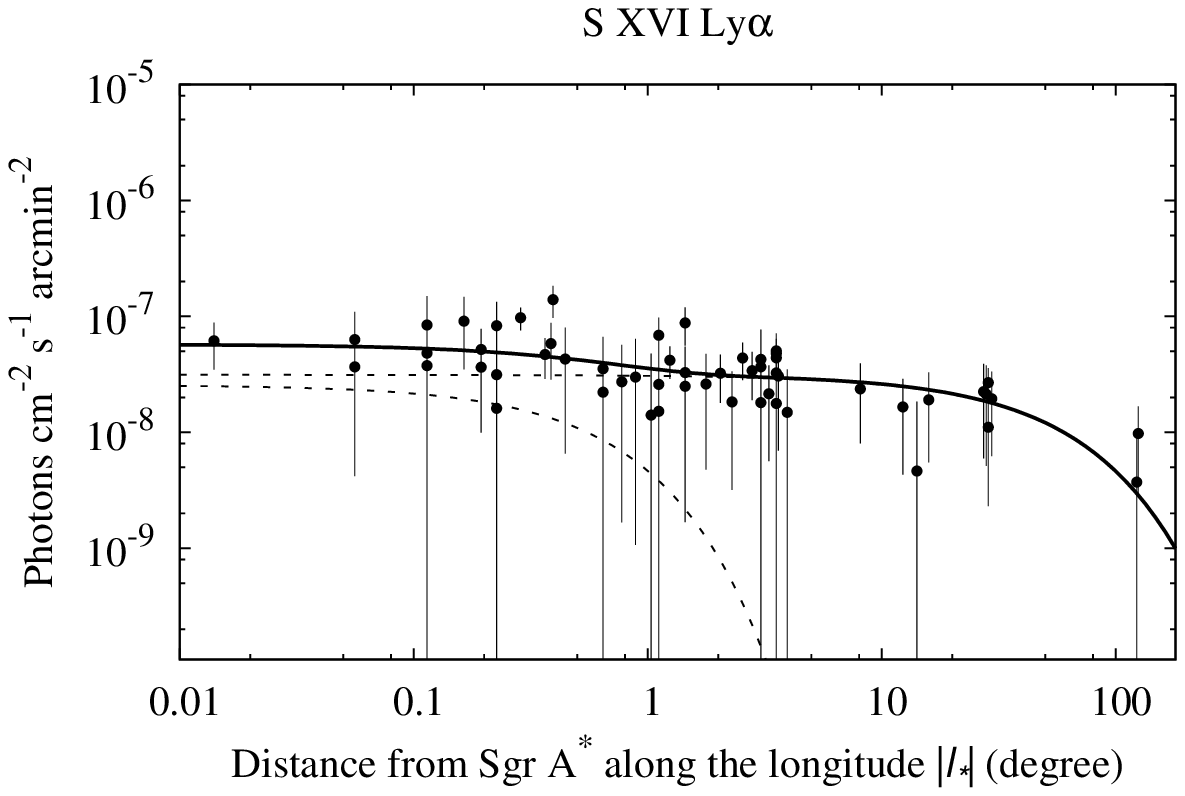}
		 \FigureFile(80mm,60mm){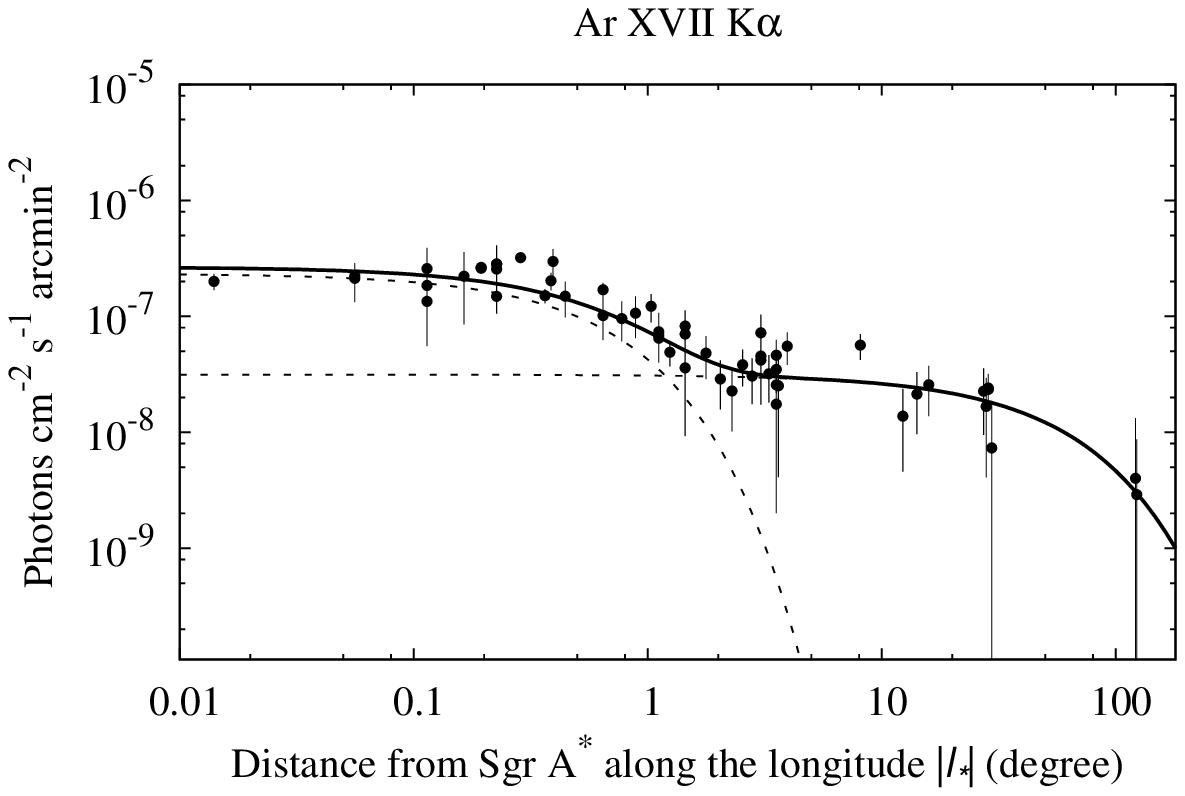}
		 \FigureFile(80mm,60mm){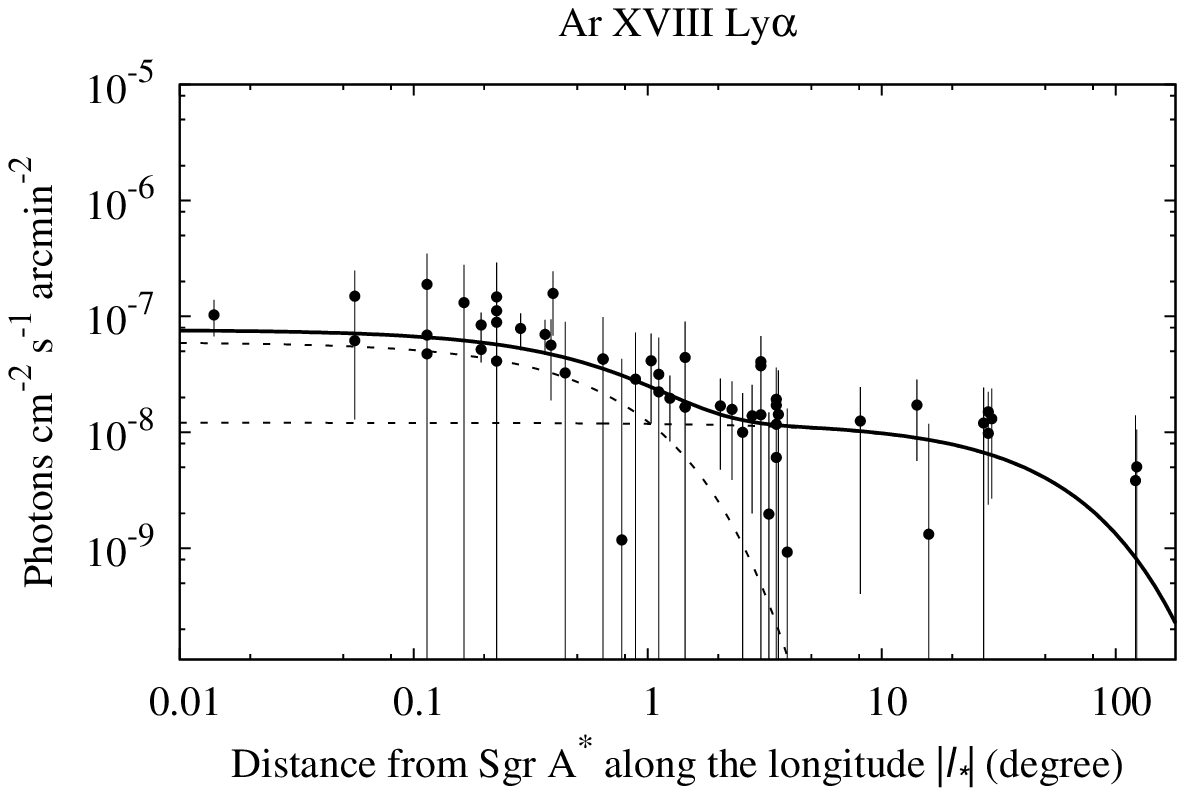}
		 \FigureFile(80mm,60mm){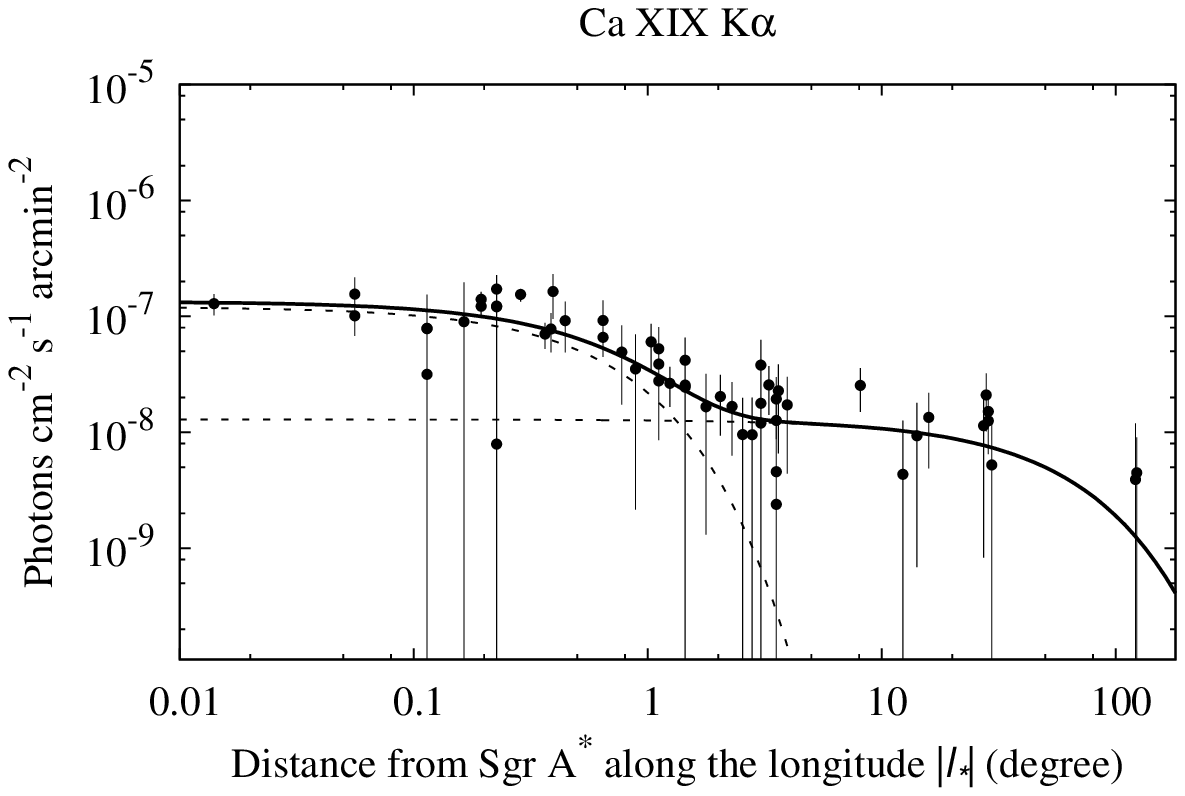}
		 \FigureFile(80mm,60mm){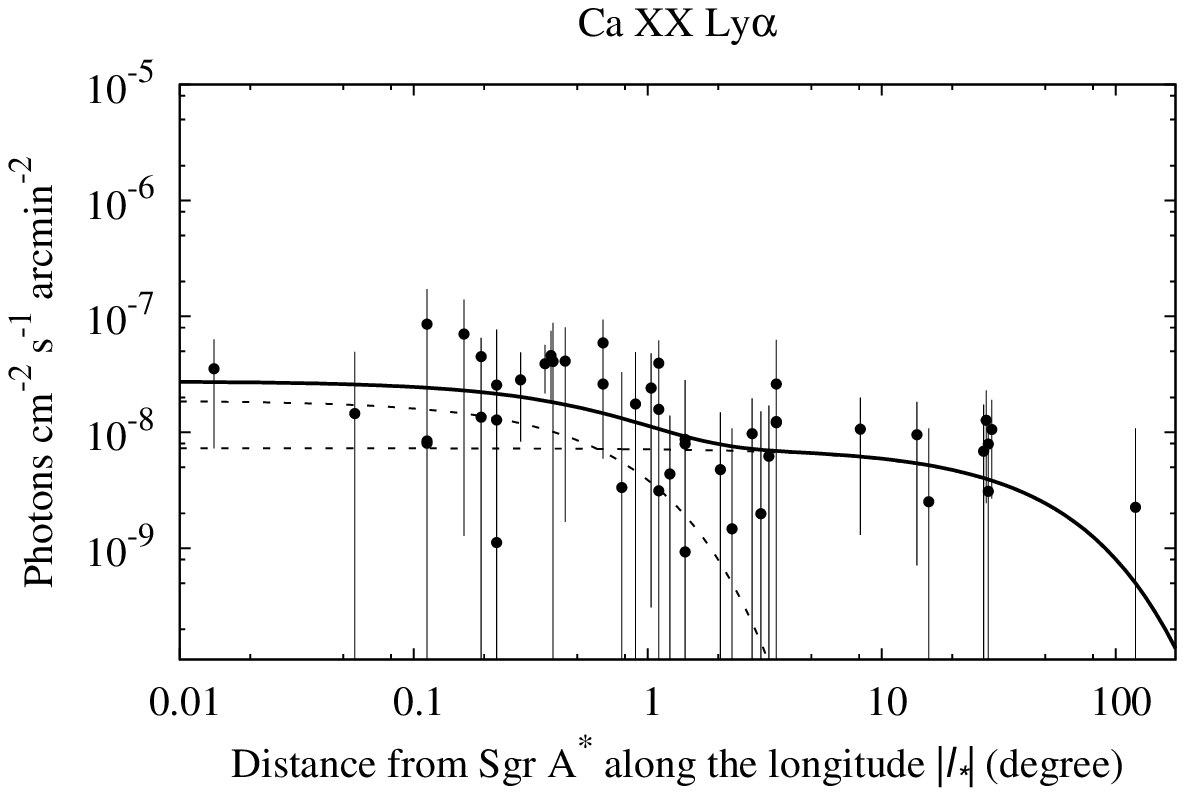}
		 \FigureFile(80mm,60mm){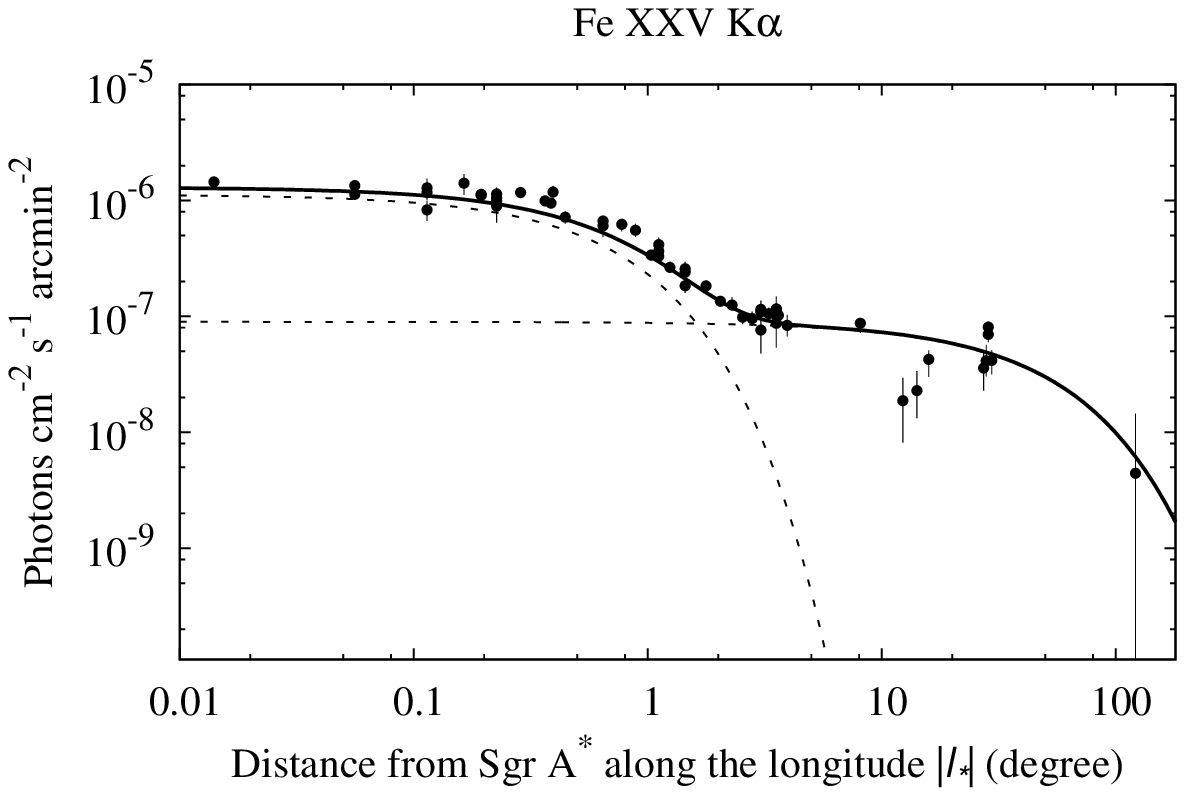}
		 \FigureFile(80mm,60mm){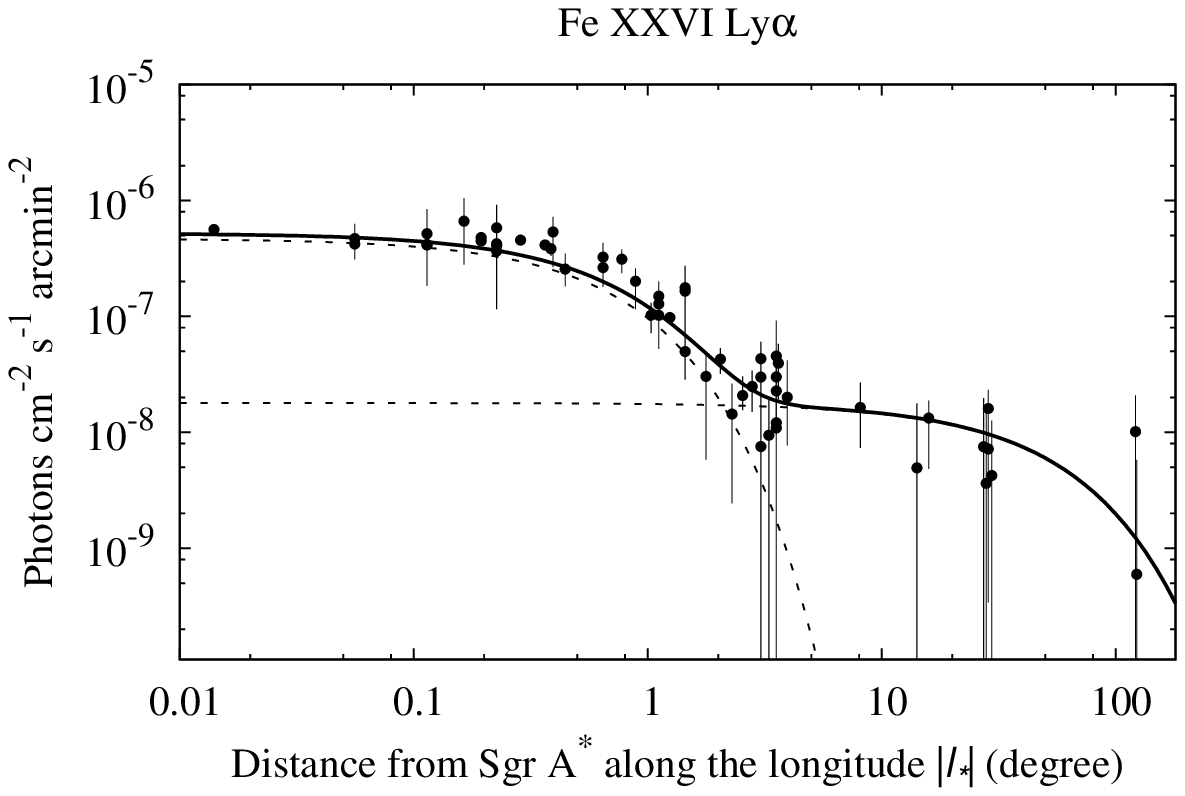}
		\end{center}
		\caption{Intensity distributions of the K-shell lines of Fe, Ca, Ar and S along the Galactic plane. }
		\label{fig:Line_Dist}
\end{figure*}

\subsection{Intensity profiles of the continuums and lines}

Using the best-fit intensities, we constructed the line intensity distributions along the Galactic plane as 
a function of angular distance from the Galactic center (Sgr A$^*$). 
In the following description, we adopted a new coordinate as 
$(l_*, b_*)=(l+\timeform{0D.056}$, $b+\timeform{0D.046}$),  
where the origin of this coordinate $(l_*, b_*)=(\timeform{0D}, \timeform{0D})$ 
is the position of Sgr A$^*$ \citep{Re04} and the $b_*=\timeform{0D}$ line 
corresponds to the Galactic plane.

The line intensity profiles were fitted with a phenomenological model given by
\begin{eqnarray}\label{eq:2exp} 
I(l_*,b_*)=I_{\rm C}\times \exp\left(-\frac{|l_*|}{l_{\rm C}}\right) \times \exp\left(-\frac{|b_*|}{b_{\rm C}}\right) \nonumber\\
+I_{\rm R}\times \exp\left(-\frac{|l_*|}{l_{\rm R}}\right) \times \exp\left(-\frac{|b_*|}{b_{\rm R}}\right),
\end{eqnarray}
where the first and the second terms of the right side of equation 1 represent the intensity profiles of the GCXE and GRXE, respectively. 
The parameters $l_{\rm C}$ and $l_{\rm R}$ are the e-folding scale lengths along the Galactic plane, while the parameters $b_{\rm C}$ and $b_{\rm R}$ are the e-folding scale heights perpendicular to the Galactic plane, for the GCXE and GRXE, respectively.

At first, we fitted the line intensity profiles with free  parameters  of $I_{\rm C}$, $I_{\rm R}$, $l_{\rm C}$, $l_{\rm R}$, $b_{\rm C}$ and $b_{\rm R}$ for all the elements independently.
Then the parameters for S \emissiontype{XV} K$\alpha$ and Fe \emissiontype{XXV} 
K$\alpha$ were well constrained with relatively small errors, while those of
the other lines had large errors. 
In detail, the scale heights $b_{\rm C}$ and $b_{\rm R}$ for S \emissiontype{XVI} Ly$\alpha$, 
Ar  \emissiontype{XVII} K$\alpha$ and  Ca \emissiontype{XIX} K$\alpha$ 
were $\sim \timeform{0D.4}$ and $\timeform{4D}$, near to the same values of S \emissiontype{XV} K$\alpha$.
Those for Ar \emissiontype{XVIII} Ly$\alpha$, Ca \emissiontype{XX} Ly$\alpha$, and Fe \emissiontype{XXVI} Ly$\alpha$ were $\sim \timeform{0D.3}$ and $\timeform{3D}$, near to  the same values of Fe \emissiontype{XXV} K$\alpha$.
We therefore linked these parameters with two groups; ``S \emissiontype{XV} K$\alpha$" and ``Fe \emissiontype{XXV} K$\alpha$" groups, as shown in table \ref{tab:profile_bestfit}. 
The spectral analyses in the next section support that the two groups roughly correspond to the low temperature plasma (LP) and high temperature plasma (HP), respectively (see figure 7).
Finally, we re-fitted the line intensity profiles simultaneously with the constraints mentioned above.  Then all the profiles were still nicely described with the two-exponential model (equation \ref{eq:2exp}) within reasonable errors of the relevant parameters.
The best-fit parameters are summarized in table \ref{tab:profile_bestfit}.

For the S and Fe lines, 
the intensity normalizations of the GCXE and GRXE ($I_{\rm C}$ and $I_{\rm R}$) were well determined. 
The ratios of Fe \emissiontype{XXVI} Ly$\alpha$ to Fe \emissiontype{XXV}  K$\alpha$ (Ly$\alpha$/K$\alpha$) are  $0.45\pm0.06$ and  $0.20\pm0.08$ in the GCXE and GRXE, respectively,  which indicate the plasma temperatures are respectively $\sim8$ and $\sim5$ keV, respectively.
On the other hand, those of S \emissiontype{XVI} Ly$\alpha$ to  S \emissiontype{XV}  K$\alpha$  are $<0.09$ and  $0.37\pm0.14$ in the GCXE and GRXE, respectively.  
Therefore the respective plasma temperatures are $kT<1 $ keV and $kT\sim$ 1.3--1.6 keV.
These facts support that the GCXE and GRXE consist of, at least, two-temperature 
plasmas as already reported by the previous works (see introduction).  

The Fe \emissiontype{I} K$\alpha$ line should be emitted by cold neutral matter and have a different origin from the above thermal lines. Therefore we fitted its intensity profile separately.      
We also fitted the intensity profiles of the 2.3--5 and 5--8 keV bands with the same phenomenological model but including the contribution of the CXB.
The spectrum and intensity of the CXB were estimated according to \citet{Ku02} with an absorption of $N_{\rm H}=6\times10^{22}$ cm$^{-2}$  (\cite{Eb01}, \cite{Ya09}).
These best-fit parameters are also listed in table \ref{tab:profile_bestfit}.

\begin{figure}[hbtp]
		\begin{center}
		 \FigureFile(80mm,60mm){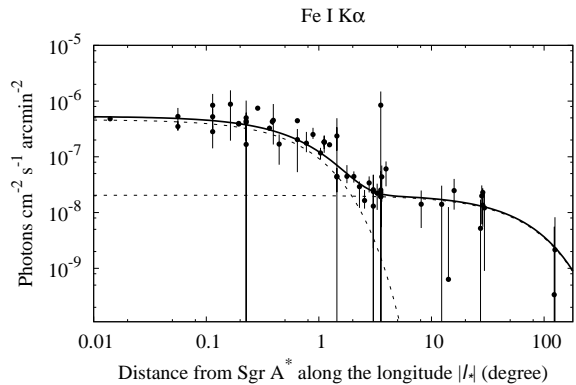}
		\end{center}
		\caption{Intensity distribution of the Fe \emissiontype{I} K$\alpha$ line  along the Galactic plane.  }
		\label{fig:FeI_Distl}
\end{figure}

\begin{figure}[hbtp]
		\begin{center}
		 \FigureFile(80mm,60mm){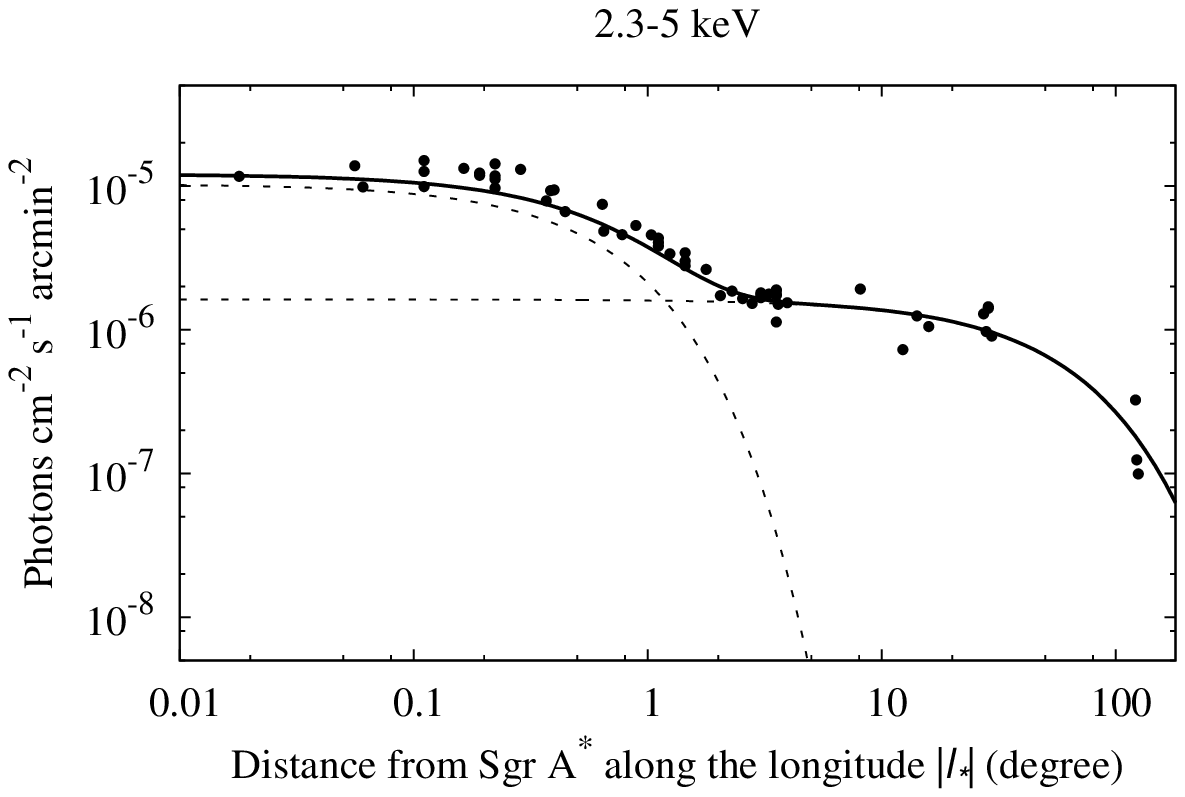}
		 \FigureFile(80mm,60mm){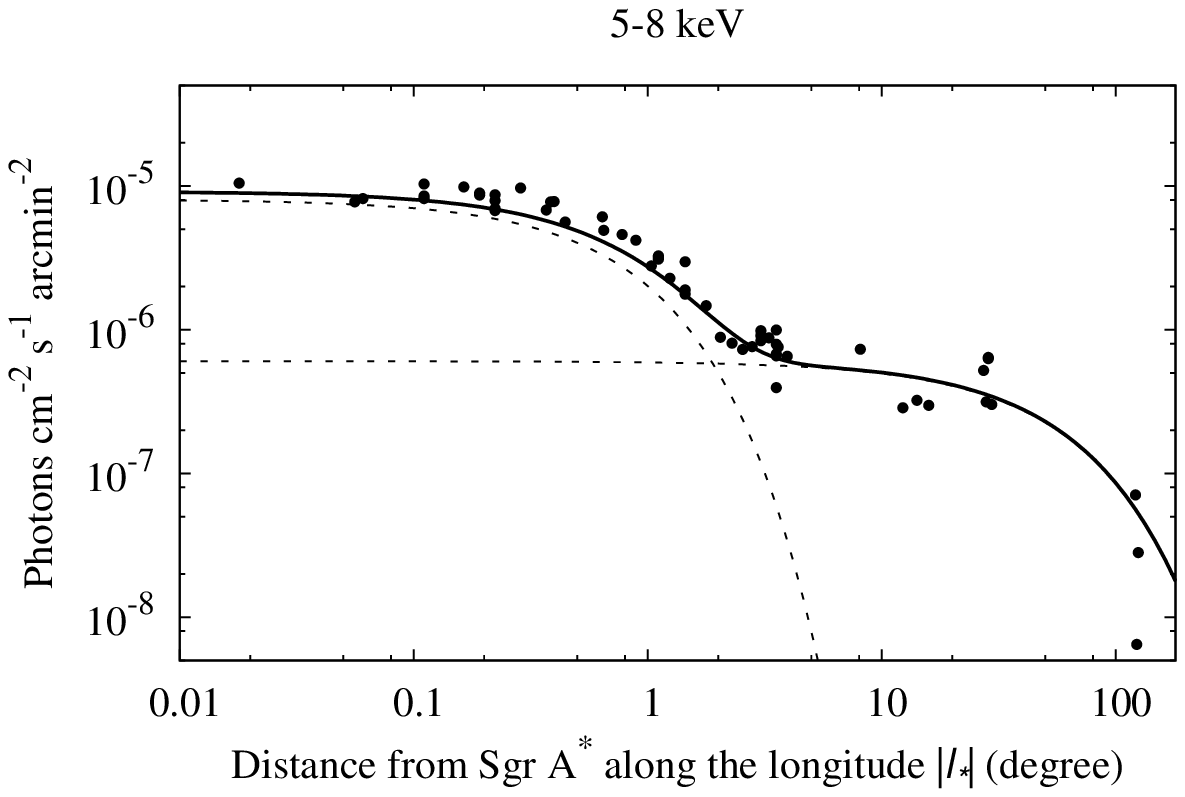}
		\end{center}
		\caption{Intensity distributions of the X-ray flues in the 2.3--5 keV (top) and 5--8 keV bands (bottom) along the Galactic plane. }
		\label{fig:Flux_Dist}
\end{figure}

\begin{table*}
\caption{Best-fit parameters of the two-exponential model.}
\label{tab:profile_bestfit}
  \begin{center}
    \begin{tabular}{cccccccc}
     \hline
      \hline
        &$I_{\rm C} {}^*$ & $I_{R} {}^*$ &$I_{\rm C}/I_{\rm R}$ &$l_{\rm C}$ (degree) &$b_{\rm C}$ (degree)  & $l_{\rm R}$ (degree) & $b_{\rm R}$ (degree)\\
\hline
S \emissiontype{XV} K$\alpha$ 	&	7.78 $\pm$ 0.47 & 0.75 $\pm$ 0.08 & 10 $\pm$ 1	&    	0.58 $\pm$ 0.05	&    0.43 $\pm$ 0.03 	&   	52 $\pm$ 13 	&    4.1 $\pm$ 1.3 \\
S \emissiontype{XVI} Ly$\alpha$ 	&	0.31 $\pm$ 0.24 & 0.32 $\pm$ 0.05 & 1.0 $\pm$ 0.8  	&	0.58$^\dagger$ &	0.43$^\dagger$ 	& 	52$^\dagger$   	& 4.1$^\dagger$ \\
Ar  \emissiontype{XVII} K$\alpha$ 	&   2.86 $\pm$ 0.24 & 0.32 $\pm$ 0.05 & 9 $\pm$ 2 &    0.58$^\dagger$ &	0.43$^\dagger$ 	& 	52$^\dagger$   	& 4.1$^\dagger$ \\
Ar  \emissiontype{XVIII} Ly$\alpha$ 	&    0.78 $\pm$ 0.21 & 0.12 $\pm$ 0.03 & 7 $\pm$ 2 &    0.63$^\ddagger$  &    0.27$^\ddagger$ &   45$^\ddagger$ &    2.9$^\ddagger$ \\
Ca \emissiontype{XIX} K$\alpha$ 	&    1.48 $\pm$ 0.19 & 0.13 $\pm$ 0.03 & 11 $\pm$ 3 &   0.58$^\dagger$ &	0.43$^\dagger$ 	& 	52$^\dagger$   	& 4.1$^\dagger$  \\
Ca \emissiontype{XX} Ly$\alpha$ 	&    0.24 $\pm$ 0.18 & 0.07 $\pm$ 0.03 & 3.4 $\pm$ 3.0 &    0.63$^\ddagger$  &    0.27$^\ddagger$ &   45$^\ddagger$ &    2.9$^\ddagger$ \\
Fe  \emissiontype{XXV} K$\alpha$ 	&    14.57 $\pm$ 0.47 & 0.91 $\pm$ 0.08 & 16 $\pm$ 2 &   0.63 $\pm$ 0.03  &    0.27 $\pm$ 0.01 &   45 $\pm$ 10 &    2.9 $\pm$ 0.5  \\
Fe \emissiontype{XXVI} Ly$\alpha$ 	&    6.08 $\pm$ 0.30 & 0.18 $\pm$ 0.04 & 34 $\pm$ 7 &      0.63$^\ddagger$  &    0.27$^\ddagger$ &   45$^\ddagger$ &    2.9$^\ddagger$ \\
Fe \emissiontype{I} K$\alpha$ 	&    6.2 $\pm$ 0.6 & 0.21 $\pm$ 0.09 & 29 $\pm$ 13 &    0.62 $\pm$ 0.09 &    0.23 $\pm$ 0.03 &   57 $\pm$ 50 &     1.1 $\pm$ 0.6  \\
\hline
2.3--5 keV 					& 	126 $\pm$ 15 	&	 16 $\pm$  1	 & 	8 $\pm$ 1 &0.63 $\pm$ 0.07  & 0.42 $\pm$ 0.04  & 59 $\pm$ 6 & 4.9 $\pm$ 0.8  \\ 
5--8 keV 						& 	101 $\pm$ 11 	& 	6.2 $\pm$ 0.6	& 16 $\pm$ 2 &0.72 $\pm$ 0.06 & 0.31 $\pm$ 0.02 & 52 $\pm$ 9 & 2.8 $\pm$ 0.5  \\
\hline
      \end{tabular}
    \end{center}
    \footnotemark[$*$] The units are 10$^{-7}$ photons s$^{-1}$ cm$^{-2}$ arcmin$^{-2}$. The interstellar absorption is not corrected.\\
     \footnotemark[$\dagger$]  These e-folding scale lengths and scale heights are linked to those of  S \emissiontype{XV} K$\alpha$. \\
         \footnotemark[$\ddagger$] These e-folding scale lengths and scale heights are linked to those of  Fe \emissiontype{XXV} K$\alpha$.
\end{table*}

The intensity distributions of S, Ar, Ca and Fe along the Galactic plane were shown in figures \ref{fig:Line_Dist} and \ref{fig:FeI_Distl}. Also the 2.3--5 keV and 5--8 keV band intensities are given in figure \ref{fig:Flux_Dist} together with the best-fit models. 
In these figures, all the data points were multiplied by factors of $I(l_*, 0)/I(l_*,b_*)$ to correct the intensity dependence along the latitude.  

From figures 1, 2 and table 1, we demonstrate a clear separation of the GCXE and GRXE. 
The GCXE is concentrated near the Galactic center (Sgr A$^*$) with  
e-folding parameters of about \timeform{0D.6}  ($l_{\rm C}$ in table \ref{tab:profile_bestfit}) 
along the longitude. The latitude extension ($b_{\rm C}$; scale height) is about 
\timeform{0D.2}--\timeform{0.D4}. 
The GRXE is more extended with the e-folding parameters of $\sim$\timeform{50D} in the longitude 
and the scale heights of  about \timeform{1D}--\timeform{5D} in the latitude .

\begin{figure}[hbtp]
		\begin{center}
		 \FigureFile(80mm,60mm){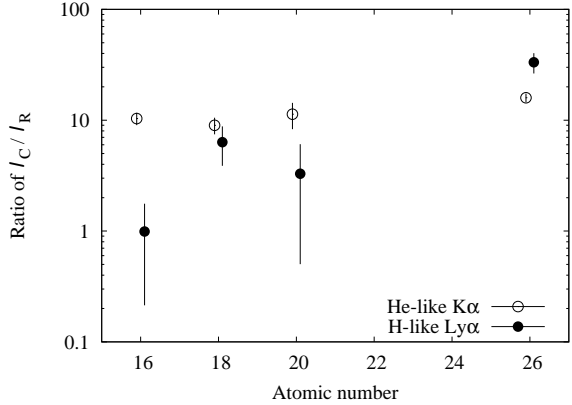}
		\end{center}
		\caption{Intensity ratio of the GCXE to GRXE ($I_{\rm C}/I_{\rm R}$) as a function of atomic numbers. 
		}
		\label{fig:I1tI2}
\end{figure}

The intensities at the center from the GCXE and GRXE are given with the parameters $I_{\rm C}$ and $I_{\rm R}$, respectively.
In order to see systematic trends of the $I_{\rm C}$/$I_{\rm R}$ ratios, we plotted the ratio in figure \ref{fig:I1tI2} as a function of atomic number. 
For the He-like K$\alpha$ lines, all the elements have roughly equal values of
$I_{\rm C}$/$I_{\rm R}$ $\sim 10$--20, while the H-like Ly$\alpha$ lines are largely scattered among the elements.
These phenomena are not due to the difference of interstellar absorption ($N_{\rm H}$) between the GCXE and GRXE, because the He-like K$\alpha$ and H-like Ly$\alpha$ lines show different trends.
The observed S line intensities may be reduced by the large $N_{\rm H}$. 
The averaged $N_{\rm H}$ of the GCXE and GRXE are  
$6\times10^{22}$ and $4\times10^{22}$ cm$^{-2}$, respectively (see section \ref{ch:sp}). 
Then the K$\alpha$ lines are absorbed by 80\% and 66\%, 
while S \emissiontype{XVI} Ly$\alpha$ lines are 70\% and 55\%, for the GCXE and GRXE, respectively.
Thus the correction factors on the ratio of $I_{\rm C}$/$I_{\rm R}$ due to the $N_{\rm H}$ absorption are 1.7 and 1.5 for the GCXE and GRXE, respectively, 
and hence give no significant change on the trends in figure 4.

A remarkable contrast is found in the profiles of the Fe \emissiontype{XXVI} Ly$\alpha$ and S \emissiontype{XVI} Ly$\alpha$ lines.
The $I_{\rm C}/I_{\rm R}$ of S \emissiontype{XVI} Ly$\alpha$ is $\sim$1 and hence  there is no clear excess of S \emissiontype{XVI} Ly$\alpha$ 
in the GCXE, while that of Fe \emissiontype{XXVI} Ly$\alpha$ is $\sim$34 times larger than that in the GRXE. 
These structures may be attributable to a complex plasma structure in the GCXE and GRXE (see discussion).

\begin{figure}[hbtp]
		\begin{center}
		 \FigureFile(80mm,60mm){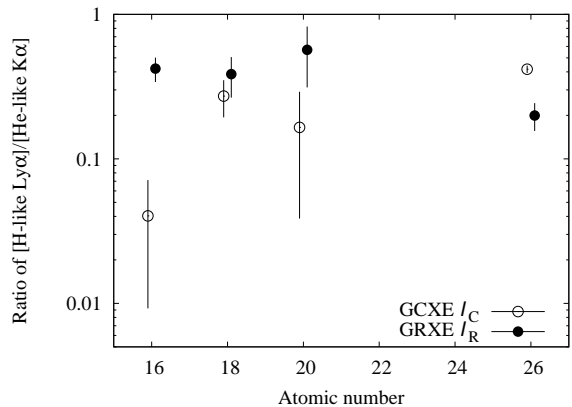}
		\end{center}
		\caption{Intensity ratio  of H-like Ly$\alpha$/He-like K$\alpha$  for the GCXE and GRXE as a function of atomic numbers.
		}
		\label{fig:HHeRatio}
\end{figure}

The complex plasma structure is also noted in the intensity ratios of H-like Ly$\alpha$/He-like K$\alpha$ for the elements in the GCXE and GRXE (figure \ref{fig:HHeRatio}). 
This figure indicates that the ionization states of S, Ar and Ca are higher in the plasma of the GRXE than those in the GCXE, while visa versa for the 
ionization state of Fe. In the next section, we present a quantitative analysis for the complexity of the GCXE and GRXE plasmas.

%%%%%%%%%%%%%%%%%%%%%%%%%%%%

\subsection{Two-temperature structures for the GCXE and GRXE}\label{ch:sp}
Since we were able to divide the GCXE and GRXE regions in the line-intensity 
distributions, mean X-ray spectra of the GCXE and GRXE were separately made 
using the data sets from the regions of ($|l|<\timeform{1D}$, 
$|b|<\timeform{0D.5}$) and ($|l|>\timeform{2D}$, $|b|<\timeform{0D.5}$), respectively. 

\begin{figure}[hbtp]
		\begin{center}
		 \FigureFile(80mm,60mm){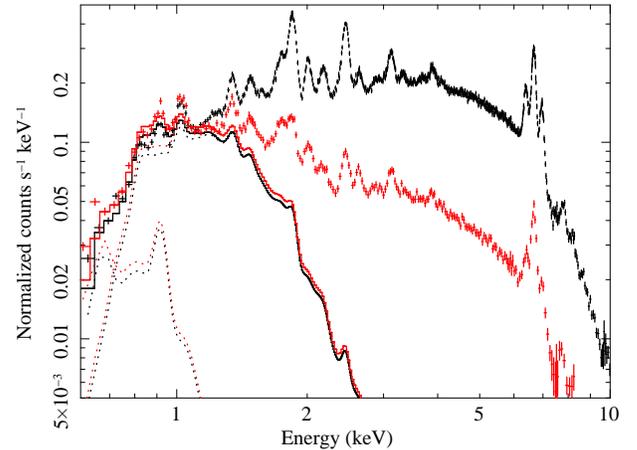}
		\end{center}
		\caption{Wide-band spectra (0.5--10 keV) of GCXE and GRXE (black and red marks, respectively) after the subtraction of the NXB. The model fitted in the 0.5--1.2 keV band is shown in the solid line.}
		\label{fig:Full_Band_w_low_model}
\end{figure}

Although we selected data regions with no-bright source by eyes (section \ref{ch:obs}), some faint sources might contaminate the selected data.  
To exclude this possibly contaminated data, we calculated standard deviations of  $(I_{\rm data}-I_{\rm model})/I_{\rm model}$, where $I_{\rm data}$ and $I_{\rm model}$ 
are the continuum intensities of the data and the predicted values by the best-fit 
two-exponential model, in the 2.3--5 keV and 5--8 keV bands (figure \ref{fig:Flux_Dist}). 
The standard deviations ($1\sigma$) are 0.34 (2.3--5 keV) and 0.24 (5--8 keV). 
We excluded the data whose intensities were over the 2$\sigma$ range 
in the either 2.3--5 keV or 5--8 keV band. 
The data used to make the mean spectra are indicated in the last column of table \ref{tab:obslog}.  
The wide-band spectra of the GCXE and GRXE in the 0.5--10 keV band are given in 
figure~\ref{fig:Full_Band_w_low_model} with the black and red marks, respectively. 
The total exposures of the spectra are 948 (GCXE) and 526 (GRXE) ks.

We see that both of the spectra have a common component below $\sim$1.2 keV. 
We fitted the 0.5--1.2 keV band spectra with two-temperature plasmas. 
The best-fit temperatures are $0.09 \pm 0.01$ and $0.59\pm0.01$~keV. 
In this paragraph, errors are estimated at the 90\% confidence ranges.
The abundance and absorption column density are 
$Z=0.05\pm0.01$ solar and $N_{\rm H}=5.6 (\pm0.5)\times10^{21}$~cm$^{-2}$, respectively.
The 0.5--1.2 keV intensities are 1.1$\times 10^{-6}$ and  9.7 $\times 10^{-7}$ ~ph.~cm$^{-2}$~s$^{-1}$~arcmin$^{-2}$
 for the  0.09 and 0.59 keV components, respectively.
 
As we found from the fitting with the phenomenological model (section 3.1), 
the intensity ratios of the He-like K$\alpha$ and H-like Ly$\alpha$ are different 
from element to element. 
The ratios are also different from the GCXE to the GRXE. 
These facts indicate that the K-shell lines from various atoms cannot be 
explained by one-temperature plasma but multi-temperature components are required.	
To investigate the origins of the GCXE and GRXE, we fitted the spectra with 
a physical model in the relevant energy range of 2.3--10 keV. 
These fits include the foreground emission with fixed intensity and shape given 
in figure \ref{fig:Full_Band_w_low_model}.

Including the Fe \emissiontype{I} K$\alpha$ (6.4 keV) line, the plasma needs at least 
three-temperature components, 
which are representative for the emission lines of the 6.4 keV (Fe \emissiontype{I} K$\alpha$, cold matter), 
the 2.45 keV (S \emissiontype{XV} K$\alpha$, low temperature plasma: LP), 
and the 6.7 keV (Fe \emissiontype{XXV} K$\alpha$, high temperature plasma: HP). 
Also the spectra include the cosmic X-ray background (CXB). 
Thus minimum requirements are that the spectra compose of the following four components;
\begin{enumerate}
\item High temperature plasma (HP) responsible for the 6.7-keV line. 
\item Low temperature plasma (LP)  responsible for the 2.45-keV line. 
\item Cold matter (CM) component that emits the 6.4-keV line. 
\item The cosmic X-ray background (CXB).
\end{enumerate}

We combined the HP and LP  and defined as the thermal plasma (TP) given by; 
\begin{eqnarray}
{\rm TP = APEC1+APEC2~[ph.~cm^{-2}~s^{-1}~arcmin^{-2}]},\label{eq:plasma}
\end{eqnarray} 
where APEC is an optically thin-thermal plasma model by \citet{Sm01}, and APEC1 and APEC2 represent the HP and LP, respectively.

The component 3 possibly originates from bombardment of X-rays, electrons, or protons on cold interstellar gas.
We designated as the cold matter component (CM) as;
\begin{eqnarray}
{\rm CM} &=& {\rm Gaussian1}  +{\rm Gaussian2} \nonumber \\
&{}&+A\times (E/{\rm keV})^{-\Gamma}~{\rm [ph.~cm^{-2}~s^{-1}~arcmin^{-2}]},\label{eq:xf}
\end{eqnarray} 
where narrow Gaussians, Gaussian1 and Gaussian2, represent Fe~\emissiontype{I} K$\alpha$ and K$\beta$ lines, respectively. 
The line center energy and intensity of Gaussian2 were linked to 1.103 and 0.125 times 
of those of Gaussian1 \citep{Ka93}. 

The cosmic X-ray background is compiled by  \citet{Ku02} as; 
\begin{eqnarray}
{\rm CXB}&=&8.2\times10^{-7} \nonumber \\
&{}&\times (E/{\rm keV})^{-1.4}~{\rm [ph.~cm^{-2}~s^{-1}~arcmin^{-2}]}. 
 \end{eqnarray}

These components are the subject of absorption, and hence our model to fit the data is;
\begin{eqnarray}
{\rm MODEL}&=&{\rm Abs1 \times(TP+ Abs2 \times CM} \nonumber \\ 
&+&{\rm Abs1\times CXB)~[ph.~cm^{-2}~s^{-1}~arcmin^{-2}]}, \label{eq:model}
\end{eqnarray}
where Abs1 is interstellar absorption toward the GCXE or GRXE, 
and Abs2 is intra-cloud absorption which emits neutral iron lines.
The Abs1 multiplies twice on the CXB component because the CXB suffers from interstellar absorption (Abs1) two times:  
those by the front and back sides of the TP component.

\citet{No10} successfully decomposed the GCXE plasma into the similar 
components of equation~\ref{eq:model}. Therefore, we applied this model 
(equation \ref{eq:model}) for the GCXE and GRXE separately. 
The temperatures of APEC1 and APEC2 were left free, but the abundances are collectively linked, keeping the ratio proportional to the solar abundance \citep{An89}.
However the Ar, Ca and Ni lines in the GCXE and the Ar line in the GRXE show excess over the models, we hence made the abundances of these elements free.

\begin{figure*}[hbtp]
		\begin{center}
		 \FigureFile(80mm,60mm){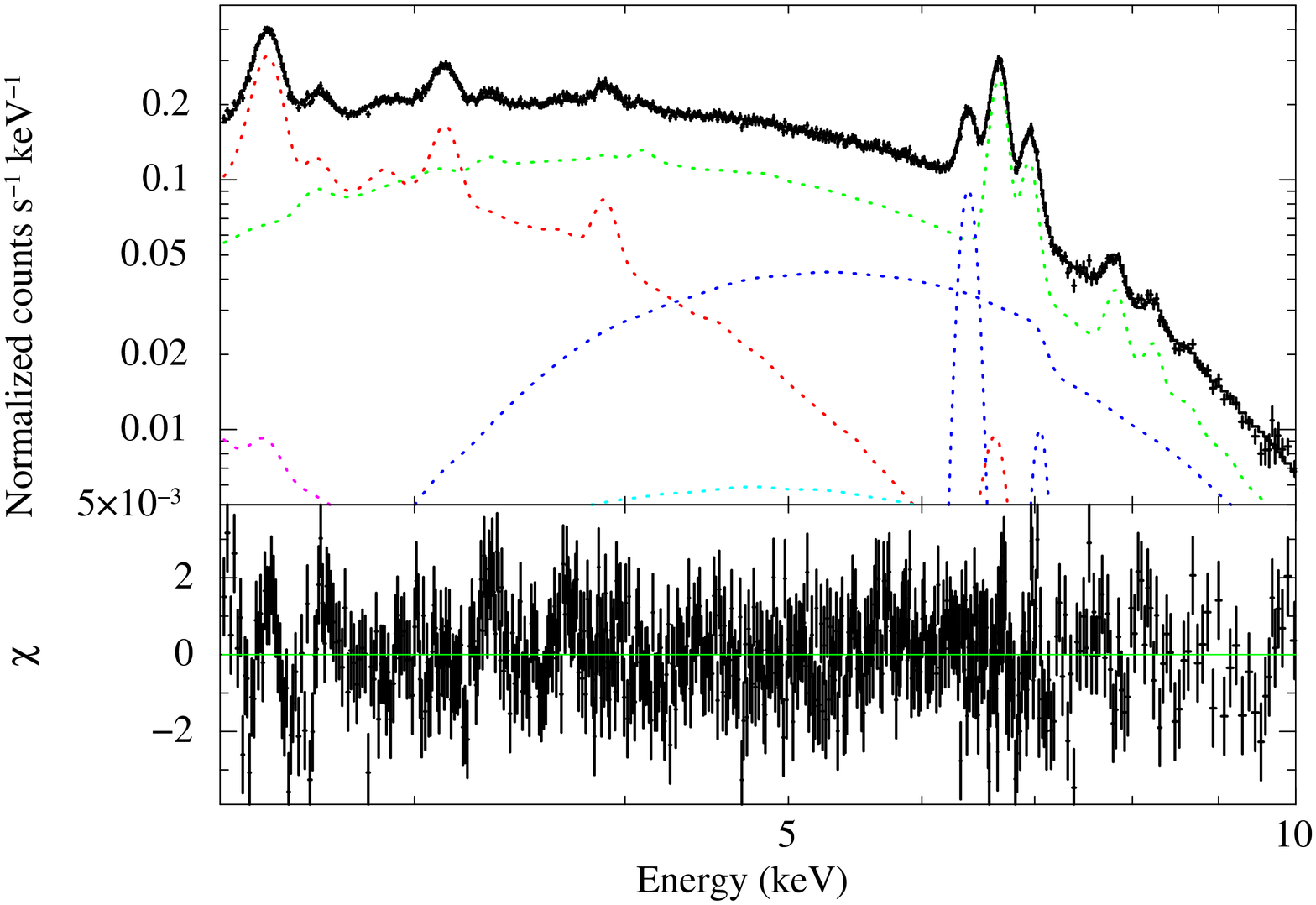}
		 \FigureFile(80mm,60mm){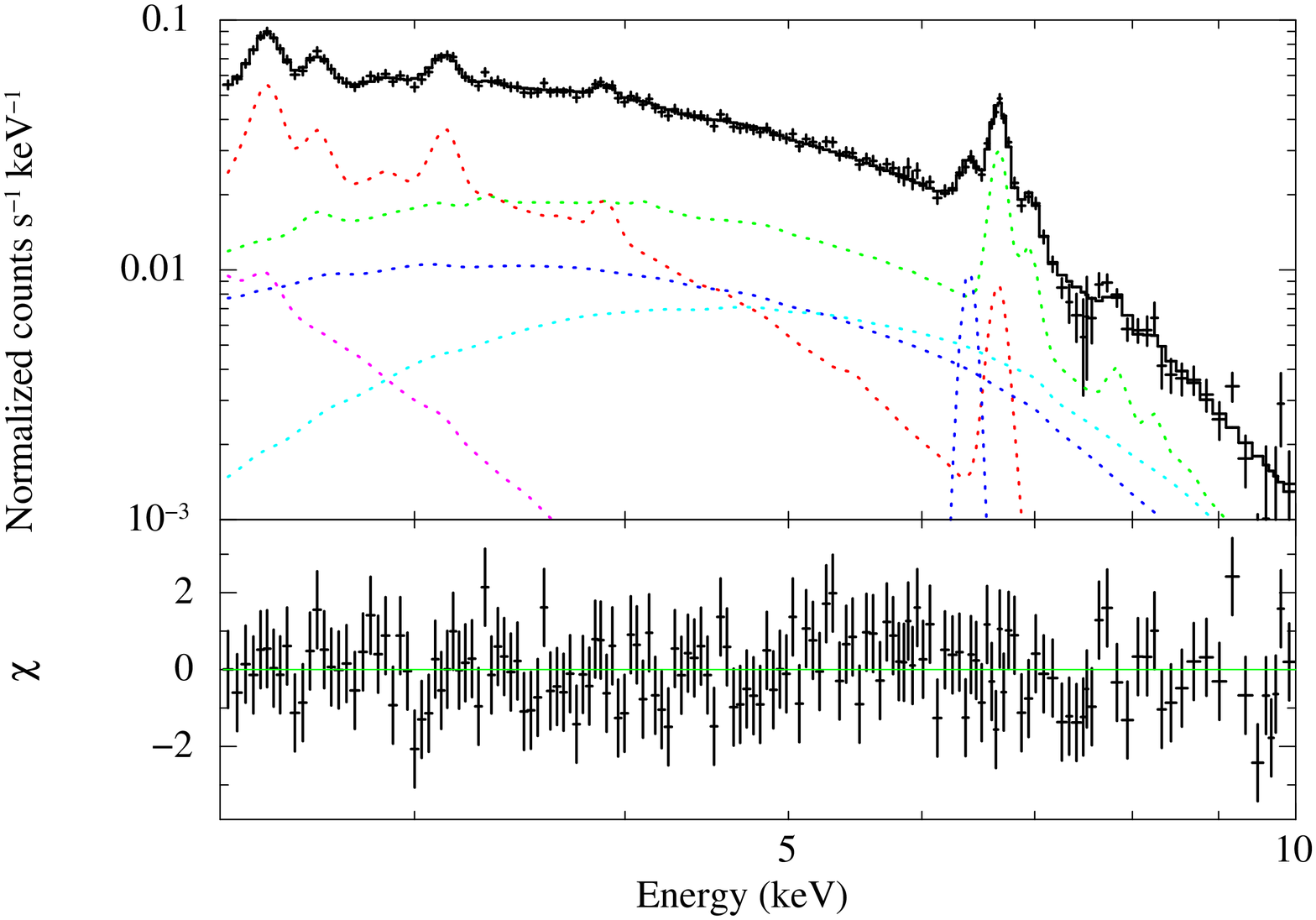}
		\end{center}
		\caption{NXB-subtracted X-ray spectra of the GCXE (left)	and GRXE (right). The solid lines are the best-fit physical model (equation \ref{eq:model}) 
		for GCXE and GRXE. The dotted lines show APEC1 (green), APEC2 (red), CM (blue), CXB (cyan) and foreground (magenta) model components, respectively.}
		\label{fig:model_fit}
\end{figure*}

For the fitting of the GRXE spectrum in the component 3, 
photon index and equivalent width (EW) of Fe~\emissiontype{I} K$\alpha$ ($EW_{6.4}$)
were fixed to the best-fit values of the GCXE. The absorption (Abs2) was fixed to be 0  
because there is no particularly dense molecular cloud 
in the GRXE region and thus we assumed that intra-cloud absorption is ignorable. 
The fitting results for the GCXE and GRXE are shown in figure~\ref{fig:model_fit}, 
and the best-fit parameters are listed in table~\ref{tab:fitting}.

\begin{table}
\caption{The best-fit physical parameters for the GCXE and GRXE$^*$.\label{tab:fitting}}
\begin{center}
\begin{tabular}{lccc}
 \hline \hline
&GCXE&GRXE\\
\hline
$N_{\rm H1}$ ($10^{22}$ cm$^{-2}$) 
&$5.59\pm0.20$  
&$4.22^{+0.34}_{-0.16}$ \\
$N_{\rm H2}$ ($10^{22}$ cm$^{-2}$) 
&$21\pm2$ 
& 0 (fixed)\\
$kT_1$ (keV) 
&$7.38^{+0.05}_{-0.10}$ 
&$6.64^{+0.40}_{-0.42}$\\
$kT_2$ (keV) 
&$0.95\pm0.02$ 
& $1.33^{+0.11}_{-0.08}$ \\
$Z$ (solar) 
&$1.25^{+0.04}_{-0.05}$ 
& $0.81^{+0.15}_{-0.09}$ \\
$Z_{\rm Ar}$ (solar) 
&$1.73^{+0.11}_{-0.15}$ 
&$1.07^{+0.21}_{-0.20}$  \\
$Z_{\rm Ca}$ (solar)
&$1.89\pm0.18$
&  0.81 (linked to $Z$) \\
$Z_{\rm Ni}$ (solar) 
&$2.38^{+0.30}_{-0.32}$
&0.81 (linked to $Z$) \\
$\Gamma$ 
&$2.13^{+0.09}_{-0.10}$
& 2.13 (fixed)\\
$EW_{6.4}$ (eV) 
&$457 \pm 29$
&  457 (fixed)\\
$EM1/EM2^\dagger$ 
&$0.24 \pm 0.02$
&$0.29 \pm 0.07$ \\
\hline
$\chi^2$/d.o.f. &755/528=1.43 &127/143=0.89\\
\hline
\end{tabular}
\end{center}
\footnotemark[$*$]  The uncertainties show the 90\% errors.\\
 \footnotemark[$\dagger$]  Emission measure of the plasma, which is proportional to $n_{\rm e} n_{\rm H} V$, where $n_{\rm e}$ and $n_{\rm H}$ are electron and hydrogen densities, respectively, while $V$ is the volume of plasma.  
\end{table}

\section{Discussion}
The intensity ratios of the He-like K$\alpha$ and 
the H-like Ly$\alpha$ are different from element to element, and 
the intensity ratios of these lines in the GCXE are different 
from the GRXE. The thermal components responsible for these lines 
are decomposed into, at least two components, the low temperature 
plasma (LP) and  high temperature plasma (HP). 
The LP temperature  in the GRXE is higher than that in the 
GCXE, while vice versa for the HP. 
These complex structures indicate that the origins of the GCXE 
and GRXE  would be a mixture 
of at least two sources with different temperatures. 
We separately discuss the origins of the HP and LP
although these may be closely related. 

\subsection{The high temperature plasma (HP)}
The most representative feature for the HP is the Fe~\emissiontype{XXV}~K$\alpha$ line.
\citet{Re09} resolved about 90\% of the Fe K$\alpha$ line at 
$(l, b)= (\timeform{0D.1}, \timeform{-1D.4})$ into many point sources. 
This position is outside of the GCXE and rather in the GRXE 
(see figure 4 in \cite{Uc11}). 
We therefore start the discussion with the assumption that the 
origin of the HP in the GRXE (not GCXE) may be the superposition 
of point sources. 
\citet{Uc11} compared the Fe \emissiontype{XXV} K$\alpha$ line distribution 
with the stellar mass distribution (SMD) and found that the GCXE shows 
3.8--19 times excess over the SMD that is normalized to the GRXE region. 
Since our observational results are essentially the same (but more extended), 
we will follow the discussion in \citet{Uc11}. 

The most probable candidate sources for the Fe \emissiontype{XXV} K$\alpha$ line are cataclysmic variables (CVs) and active binaries (ABs) (e.g. \cite{Yua12}), because their spectra are represented by a high-temperature plasma with large equivalent widths of the 6.7 keV  (Fe \emissiontype{XXV}  K$\alpha$; hereafter, $EW_{6.7}$) and 6.97 keV lines (Fe \emissiontype{XXVI}  Ly$\alpha$; $EW_{6.9}$), associated with the 6.4 keV (Fe \emissiontype{I}  K$\alpha$; $EW_{6.4}$) line, which are similar to those of the GRXE. The mean $EW_{6.4}$, $EW_{6.7}$ and $EW_{6.9}$ of the X-ray bright CVs in the luminosity range of $10^{30-33}$ erg s$^{-1}$ are $\sim$100 eV, $\sim$200  eV and $\sim$100 eV, respectively  (Ezuka \& Ishida 1999, Rana et al. 2006). 
We further analyzed the Suzaku archive of ABs and found that their mean $EW_{6.4}$, $EW_{6.7}$ and $EW_{6.9}$ are $\sim60$ eV, $\sim$560 eV and  $\sim$80 eV, respectively.
On the other hand, our result (figure \ref{fig:model_fit}) shows that the mean  $EW_{6.4}$, $EW_{6.7}$ and $EW_{6.9}$ in  the GRXE are $\sim$110 eV,  $\sim$490 eV and $\sim$110 eV, respectively. 
Thus, any mixture of the bright X-ray CVs and ABs cannot explain $EW_{6.7}$,   $EW_{6.4}$ and $EW_{6.9}$ of the GRXE simultaneously. In order to explain by unresolved point sources, we need faint sources with lager equivalent widths  of iron than those of the well-known CVs and ABs.

Unlike the GRXE, the majority of the GCXE has not been resolved into the point sources (\cite{Mu04}, \cite{Re07GC}).  Furthermore, the line intensity ratio of Fe \emissiontype{XXVI} Ly$\alpha$/Fe \emissiontype{XXV} K$\alpha$  is significantly larger in the GCXE than in the GRXE (see figure \ref{fig:HHeRatio}).     
Abundances are also different, 1.25 and 0.81 of the solar for the GCXE and GRXE, respectively. Therefore the GCXE is difficult to be explained by the same kind of point sources in the GRXE. We need different type of point sources with higher temperatures and abundances. This may be too artificial, and hence point source origin for the HP in the GCXE may not be favored.  

One possibility is that a significant fraction of the HP would come from diffuse optically thin thermal plasma as already proposed by \citet{Ko07b} although small fraction of the HP in the GCXE is due to the same point source origin as that in the GRXE.  
The HP in the GCXE has the scale height of $\sim$40 pc (\timeform{0D.27}) and the temperature of $kT\sim$7 keV (sound velocity $\sim$ 1500 km s$^{-1}$), and hence the dynamical age is $\sim2\times10^4$ yr. 
The luminosity of the GCXE (5--8 keV) is 
$6 \times 10^{35}$ erg s$^{-1}$, and 
the electron density is estimated to be $\sim$0.05 cc$^{-1}$. 
Hence the thermal energy of the HP is  $1\times10^{53}$ erg.  If the diffuse plasma is made by multiple supernova (SN) explosions, the rate of SNe is estimated to be higher than $5\times10^{-3}$ yr$^{-1}$. 
Since the mass in the GCXE HP region  ($\sim$40 pc $\times$ 90 pc) is $\sim$1\% of the total mass of the Galaxy (e.g. \cite{Me96}, \cite{La02}), the SNe rate of $5\times10^{-3}$ yr$^{-1}$ is unreasonably  high (see \cite{Di06} and their supplement). 
Furthermore, the temperatures of supernova remnants (SNRs) of the age of  $\sim 10^4$ yr is significantly lower than that of the HP (7 keV) in the GCXE, and hence the multiple-SN scenario would be unlikely.

Other possibility is that the HP in the GCXE may be made by violent flares of Sgr A$^*$ as suggested by  \citet{Ko96}. To produce such a high temperature ($\sim$7 keV) and large scale ($\sim$40 pc $\times$ 90 pc) plasma, 
many big flares would be necessary within the past $10^4$ yr. The most recent big flare may be that in some 100 years ago at Sgr A$^*$, which is well established by time variability of the 6.4 keV line from molecular clouds (\cite{Mu07}, \cite{Ko08}, \cite{In09}, \cite{Ko09}, \cite{Po10}, \cite{No11}, \cite{Ca11}, \cite{Ca12}).
 Looking back to the older ages, the recent discovery of the ``Fermi Bubble''  \citep{Su10} may be responsible to some fraction of the HP. 

\subsection{The low temperature plasma (LP)}
The low temperature plasma (LP) in the GRXE has been already reported by \citet{Ka97}. 
Our results are based on more extended and high quality data.  
The spatial distribution (scale length) of the LP is almost the same as the HP, but the scale height is larger than the HP.  \citet{Ka97} also reported a larger scale height of LP than that of HP, however their analysis did not separate the foreground emission of  $kT=0.6$ keV from the LP (see figure \ref{fig:Full_Band_w_low_model}) , thus their scale height of the LP may be a subject of significant errors.
Similar but slightly different LP is also found from the GCXE with the scale height is significantly larger than that of the HP

Using the best-fit distribution
and plasma parameters, we estimated that the luminosity (2.3--5 keV), density and total energy of the LP in the GCXE  are $5\times10^{35}$ erg s$^{-1}$, 0.07 cc$^{-1}$ and $1\times10^{52}$ erg, respectively.
In the GRXE, they are
$6\times10^{37}$ erg s$^{-1}$, $2\times10^{-3}$ cc$^{-1}$ and $5\times10^{55}$ erg, respectively.
From the LP temperature of $kT \sim$1 keV (sound velocity $\sim$600 km s$^{-1}$) and the scale heights of $\sim55$ pc ($b_{\rm C} \sim\timeform{0D.4}$) and 550 pc ($b_{\rm R}\sim\timeform{4D}$), the dynamical ages are estimated to be $9\times10^4$ yr and $9\times10^5$ yr, for the GCXE and GRXE, respectively. 

The temperature of $kT\sim$1 keV is typically found in young and intermediate aged SNRs, 
and hence it may be conceivable that the plasma is due to multiple SNe within the past $10^5$ yr.	
To explain the total energy of the LP, the SN rate of $10^{-4}$ yr$^{-1}$ and 0.5 yr$^{-1}$ for the GCXE and GRXE would be required, respectively. 
For the GCXE, this number is not unreasonable.
For the GRXE, the multiple-SN scenario alone is difficult to explain its origin. 
However, unlike the Fe K$\alpha$ line, \citet{Re09} resolved less than 50\% of
the low energy band of the GRXE to point sources (see figure 2 in their paper), and hence the point source fraction in the LP in the GRXE would be smaller than those in the HP.
Still some fraction of the LP in the GRXE would be due to point sources.	
Most probable candidate	would be dwarf M stars (dMs), because these exhibit coronal activities with $\sim$1 keV plasma and the number density is very high (\cite{Ma09}). 

\subsection{Fe \emissiontype{I} K$\alpha$ emissions}
This paper mainly targets highly ionized lines but we also comment about Fe \emissiontype{I} K$\alpha$ line emissions a little.
Clumpy and bright Fe \emissiontype{I} K$\alpha$ line emissions in the Galactic center region have been studied well since the ASCA satellite (e.g. \cite{Ko96}).
Their origin is thought to be X-ray reflection nebulae (XRNe) irradiated by the past X-ray flare of Sgr A$^*$ due to their large EWs ($EW_{6.4}\sim1$ keV) and time variability (e.g. \cite{KoIn07}, \cite{In09}).
In this paper, we excluded these previously reported XRN regions from the GCXE spectra, and the EW to the associated continuum ($EW_{6.4}\sim460$ eV) is smaller than that of the XRN.
The origin of the faint Fe \emissiontype{I} K$\alpha$ emission in the GCXE might be different  from that of the XRNe.

The origin of the Fe \emissiontype{I} K$\alpha$ line in the GRXE is also not clear.
One may argue that superpositions of CVs is one possibility, because CVs emit not only the Fe \emissiontype{XXV} K$\alpha$ but also Fe \emissiontype{I} K$\alpha$ lines. 
However, both the Fe \emissiontype{XXV} and Fe \emissiontype{I} K$\alpha$ lines are hardly to be the
same origin, because the scale height of  the Fe \emissiontype{I}  K$\alpha$ line in the GRXE region
is significantly smaller than that of the Fe \emissiontype{XXV}  K$\alpha$ line (see table \ref{tab:profile_bestfit}).

The relatively smaller EW of $EW_{6.4}\sim460$ eV than that of XRNe may prefer the scenario that the origin is due to the interaction between high-energy electrons and interstellar neutral matter (e.g. \cite{Yu12}).  More quantitative details are  beyond the scope of this paper.

\section{Summery}
We summarize the results of this paper as;
\begin{enumerate}
\item We have analyzed all the Suzaku archive in the Galactic plane within $|b|<\timeform{5D}$. 
\item We presented a global structure of the Galactic plane in the K-shell line intensities from highly
ionized S, Ar, Ca and Fe as well as neutral Fe \emissiontype{I} and the continuum band in 2.3--5 and 5--8 keV. 
\item We clearly separated the Galactic center X-ray emission (GCXE) from the Galactic ridge X-ray emission (GRXE) with the scale length (in the longitude) of $\sim\timeform{0D.6}$ and $\sim\timeform{50D}$, and the scale heights (in the latitude) of $\sim\timeform{0D.2}$--$\timeform{0D.4}$ and $\sim\timeform{1D}$--$\timeform{5D}$, respectively. 
\item The K-shell line intensity ratios of H-like and He-like ions (Ly$\alpha$/K$\alpha$) suggest that the GCXE and GRXE plasmas have multi-temperature structures.
\item The Ly$\alpha$/K$\alpha$ profiles along the Galactic plane are different from element to element. 
\item The most striking contrast of the Ly$\alpha$ profiles is found in the S and Fe atoms. The former
shows almost no excess while the latter exhibits pronounced excess in the GCXE. 
\item The spectra of the GCXE and GRXE are decomposed into two temperature plasmas of about
1 and 7 keV. The temperature of the 1-keV plasma in the GCXE is lower than that in GRXE, and vice versa for the 7-keV plasma.

\end{enumerate}

\bigskip
The authors thank all of the Suzaku team members, especially T. Yuasa, and  K. Makishima for their comments and useful information.
 We also thank V. Dogiel for improving the paper.
 HU is supported by Japan Society for the Promotion of Science (JSPS) Research Fellowship for Young Scientists.
KK is supported by JSPS KAKENHI Grand Numbers 23000004 and 24540229.
MN is supported by JSPS KAKENHI Grand for the young scientist (B), Number 24740123.
This work was supported by the Grant-in-Aid for the Global COE Program ``The Next Generation of Physics, Spun from Universality and Emergence" from the Ministry of Education, Culture, Sports, Science and Technology (MEXT) of Japan. This work was also supported by Grant-in-Aid of the JSPS Nos. 10243007 and 23340047.

\end{document}